\documentclass[10pt,aps,prl,twocolumn,superscriptaddress,groupedaddress,nofootnoteinbib]{revtex4-1}
\pdfoutput=1 
\usepackage{fancyhdr}
\usepackage{graphicx}
\usepackage{xspace}
\usepackage{rotating}
\usepackage[normalem]{ulem}
\usepackage{braket}
\usepackage{verbatim}
\usepackage{xcolor}
\usepackage{hyperref}
\usepackage[utf8]{inputenc}
\graphicspath{{./}{./Figs/}{./figs/}{../Figs/}}

\usepackage{amsfonts}
\usepackage{mathtools}
\usepackage{pifont}

\newcommand{\benum}{\begin{enumerate}}
\newcommand{\eenum}{\end{enumerate}}
\newcommand{\bi}{\begin{itemize}}
\newcommand{\ei}{\end{itemize}}

\newcommand{\be}{\begin{equation}}
\newcommand{\ee}{\end{equation}}

\newcommand{\bea}{\begin{eqnarray}}
\newcommand{\eea}{\end{eqnarray}}
\newcommand{\beq}{\begin{eqnarray}}
\newcommand{\eeq}{\end{eqnarray}}

\newcommand{\Rmnum}[1]{\expandafter\@slowromancap\romannumeral #1@}

\DeclareMathOperator\erfc{erfc}

\begin{document} 
\title{
Detectable Gravitational Wave Signals from Inflationary Preheating}

\author{Yanou Cui}
\email[]{yanou.cui@ucr.edu}
\affiliation{Department of Physics and Astronomy, University of California, Riverside, CA 92521, USA}
\author{Evangelos I. Sfakianakis}
\email[]{esfakianakis@ifae.es}
\affiliation{
Institut de Fisica d'Altes Energies (IFAE),
The Barcelona Institute of Science and Technology (BIST),
Campus UAB, 08193 Bellaterra, Barcelona
}

\date{\today}

\begin{abstract}
We consider gravitational wave (GW) production during preheating in hybrid inflation models where an axion-like waterfall field couples to Abelian gauge fields. Based on a linear analysis, we find that the GW signal from such models can be within the reach of a variety of foreseeable GW experiments such as LISA, AEDGE, ET and CE, and is close to that of LIGO A+, both in terms of frequency range and signal strength.
Furthermore, the resultant GW signal is helically polarized and thus may distinguish itself from other sources of stochastic GW background. Finally, such models can produce primordial black holes that can compose dark matter and lead to merger events detectable by GW detectors.
\end{abstract}
\maketitle
\noindent\textbf{Introduction.}
Inflation is an early era of exponential expansion of the Universe, postulated as a leading solution to the flatness and horizon problem as well as a source of the observed cosmic structure today \cite{Guth:2005zr, Linde:1983gd, Planck:2018jri}. A transitional period of reheating is necessary to bridge the extremely cold and empty Universe with the onset of the hot Big Bang \cite{Albrecht:1982mp}. The reheating epoch can be very violent, marked by a phase of preheating, when rapid particle production occurs through parametric resonances excited by the inflaton field \cite{Kofman:1997yn, Kofman:1994rk, Amin:2014eta}. The large, time-dependent field inhomogeneities involved makes preheating a natural, well-motivated source for GW production. 

Stochastic GW background (SGWB) signals from preheating have been investigated in a variety of inflationary settings, including gauge preheating after axion inflation  \cite{Adshead:2019lbr, Adshead:2018doq}, self-resonance after single-field inflation and oscillon formation \cite{Lozanov:2019ylm, Antusch:2016con, Amin:2018xfe, Hiramatsu:2020obh} and tachyonic preheating from a waterfall transition \cite{Garcia-Bellido:2007nns, Garcia-Bellido:2007fiu} (\cite{Dufaux:2010cf} also includes a $U(1)$ charge for the waterfall fields).
Unfortunately, the current lore based on existing literature is that the GW signal from preheating is generally beyond the reach of foreseeable GW experiments: either the frequency of the resultant GW signal is too high to be covered, or the amplitude of the signal is too small to be detectable with near future GW detectors
\footnote{The scenario in \cite{Garcia-Bellido:2007nns, Garcia-Bellido:2007fiu} can lead to GW signal of $\Omega_{\rm GW}\sim10^{-15}$ within the frequency range of the relatively futuristic BBO experiment, yet much weaker than the typical prediction in our model, while the one in \cite{Dufaux:2010cf} can potentially lead to a detectable secondary peak in the advanced LIGO band.
}.

In this {\it Letter} we demonstrate a promising GW detection prospect based on a preheating scenario in the framework of hybrid inflation, where a prolonged waterfall phase allows for an efficient transfer of energy from the scalar sector to an Abelian gauge field, coupled to the waterfall fields via a Chern-Simons (CS) term. This simple model can lead to a detectable, helically polarized SGWB signal, with an optimal peak amplitude of $\Omega_{\rm GW}\sim 10^{-10}$ and the peak frequency over the wide range of $\sim(10^{-9},100)$ Hz or beyond,
depending on the Hubble scale during inflation. 
(The lowest frequency $\sim10^{-9}$ Hz results from the smallest inflationary Hubble scale allowing for a reheating temperature above $\sim$ MeV, thus compatible with  Big Bang Nucleosynthesis (BBN) constraints.) 
Such signals are thus observable with both near future and more futuristic experiments such as LISA, SKA, Einstein Telescope (ET), Cosmic Explorer (CE), AEDGE, TianGO, DECIGO, BBO, while being within an order of magnitude of the projected LIGO A+ sensitivity, and possibly with recently proposed $\mu$-Hz range GW detectors,  \cite{Binetruy:2012ze, EinsteinTelescope, CosmicExplorer, Bertoldi:2019tck, Yagi:2011wg, Sato_2017,Janssen:2014dka,TianGO, Blas:2021mqw,Sesana:2019vho}.

We took a simplified approach by solving linear evolution equations and neglecting back-reaction which can be justified as a reliable approximation with the benchmark parameters we consider. Dedicated simulation is desirable to fully exploit this promising scenario which warrants further investigations.

The scenario we consider also generally leads to production of primordial black holes that may contribute to relic dark matter and results in merger events observable by GW experiments, which complements the SGWB signals from the early Universe preheating.

\medskip

\noindent\textbf{Model.}
We consider a model consisting of a real scalar field $\psi$ and a scalar field multiplet $\phi_i$ with $i=1,2,...,\mathcal{N}$, as well as an Abelian gauge field $A_\mu$ with $F_{\mu\nu} =\partial_\mu A_\nu -\partial_\nu A_\mu$. The action is 
\beq
\label{eq:action}
\begin{split}
S = \int d^4 x \sqrt{-g} \left [
{M_{\rm Pl}^2 \over 2}R 
- \sum_i{1\over 2} \partial_\mu   \phi_i  \partial^\mu  \phi_i 
\right .
- {1\over 2} \partial_\mu \psi \partial^\mu \psi
\\
\left . - V(\psi,\phi_i) -{1\over 4}F_{\mu\nu}F^{\mu\nu}
- {1\over 4f} \sum_i  {\phi_i\over \Lambda_i} F_{\mu\nu}\tilde F^{\mu\nu}
\right ]
\end{split}
\eeq
where $M_{\rm Pl}$ is the reduced Planck mass, and the two-field scalar potential is 
\beq
\begin{split}
V(\psi,\phi_i)& =
\\
V_0 &+ V_1(\psi) - {m_0^2(\phi_i)^2\over 2}\left (1 - {\psi^2\over \psi_0^2}\right ) + {g\over 4 }(\phi_i)^4 \, ,
\end{split}
\eeq
where $(\phi_i)^2\equiv \Sigma_i \phi_i\phi_i$, indicating an $O({\cal N})$ symmetry,
and we used the mostly plus metric convention $(-,+,+,+)$. $\psi$ and $\phi_i$ play the role of the inflaton and the waterfall fields respectively, as in the hybrid inflation scenario \cite{Linde:1993cn, Linde:1991km, Garcia-Bellido:1996mdl, Linde:1997sj, Randall:1995dj}.
With ${\cal N}\geq4$ the potentially problematic formation of stable topological defects can be avoided. We choose ${\cal N}=4$ as a benchmark.
Note that the Chern-Simons couplings  break the ${\cal O}({\cal N})$ symmetry, indicating that the topological defects may be unstable and innocuous to cosmology, even for ${\cal N}<4$. We leave a detailed analysis of this possibility for future work.
While ${\cal N}$ does not affect GW production, as it can be offset by a corresponding change in $\Lambda_i$, it has an important effect on primordial black hole (PBH) production \cite{Halpern:2014mca}.\\
 \indent   During inflation $V(\psi,\phi_i)$ is dominated by the constant term, $V_0 \gg V_1(\psi)$, and thus the energy scale of inflation can be well approximated as
$\rho\simeq V_0$ and the Hubble scale as
$H^2\simeq {V_0 /( 3M_{\rm Pl}^2)}$ (A constant Hubble scale is an excellent approximation since $V/V_0\sim1- {\cal O}(0.01)$ even after the waterfall transition, up to one $e$-fold before the end of inflation). As a result, the slow-roll parameter 
$\epsilon\simeq {M_{\rm Pl}^2\over 2} (\partial_\psi V/V)^2 \simeq 
{M_{\rm Pl}^2\over 2} (\partial_\psi V_1/V_0)^2\ll 1
$. This in turn leads to a small tensor to scalar ratio
 $r=16\epsilon\ll1$. 

The potential of the inflaton (or timer) field $V_1(\psi)$ is responsible for producing the density fluctuations and must be compatible with the observed value of $n_s$.
While a simple quadratic potential leads to a blue spectral tilt, adding higher order terms with $\partial^2_{\psi}V_{1}<0$ during CMB horizon exit can rectify this (see also \cite{Clesse:2015wea, Halpern:2014mca}).
 The classical value of the waterfall fields $\phi_i$ is driven to $\phi_i=0$ for $\psi>\psi_0$, while the fields become tachyonic for $\psi<\psi_0$.  The term $(g/ 4)(\phi_i)^4$ ensures the positivity of  the potential, $V(\psi,\phi_i)\ge0$. 
 As discussed in the next Section, our results are largely insensitive to the concrete choice of $V_1(\psi)$, provided that it can be approximated by a quadratic term near the waterfall transition.

In order for the waterfall transition to occur, the condition $m_0>H$  must be satisfied (see \cite{Halpern:2014mca} for the relation of this inequality to ${\cal N}$). 
The Chern-Simons coupling between $\phi_i$ and $A_\mu$ is responsible for transferring the energy from the scalar sector to relativistic degrees of freedom via preheating. Assuming an instantaneous reheating
 which is realized in our benchmark examples, the reheating temperature is
$T_{\rm RH}\sim \sqrt{M_{\rm Pl} H} = \sqrt{M_{\rm Pl} H} \sim V_0^{1/4}$, which must be above the temperature where BBN occurs, $T_{\rm BBN}\sim$~MeV. This provides a constraint on $H$ values that we will consider. 

\medskip


\noindent\textbf{Waterfall Field Dynamics.}
At the onset of the waterfall transition, $\phi_i$ can be described by quantum fluctuations in de-Sitter space, with $\langle \phi_{i,{\rm init}}^2\rangle \sim H^2/4\pi^2$.
At the end of the waterfall phase, the full potential $V(\psi,\phi_i)$ must relax to zero. Assuming that at the time ($t_{\rm end}$) $V_1\to 0$, this leads to
$
V_0 -{1\over 2}m_0^2 \phi_{\rm end}^2 +{g\over 4}\phi_{\rm end}^4 =0
$, where $\phi_{\rm end}^2 \equiv (\phi_i)^2$ at $t_{\rm end}$, and consequently, 
$
\phi_{i, \rm end} = {m_0/ \sqrt{g}}$ and $g = {m_0^4/ 4V_0}$. Using the estimates for various components, such as $V_0\simeq 3H^2 M_{\rm Pl}^2$ and $m_0\sim O(H)$, we find
$g\sim (H/M_{\rm Pl})^2\ll1$ and
$\phi_{\rm end} \sim M_{\rm Pl}$.
 We can see that $\phi_i$ would transverse a very large distance, from $|\phi_{i,{\rm init}}|\sim H$ to $\phi_{i,{\rm end}} \sim M_{\rm Pl} $ during the waterfall phase. 

\bigskip

Our justified assumption of a constant Hubble scale simplifies the calculations and leads to a trivial relation between cosmic time and $e$-folding number, $N=H t$. 
The inflaton's equation of motion, neglecting the back-reaction from the waterfall field, is
\beq
\ddot \psi  + 3 H \dot \phi + m_\psi^2 \psi=0
\eeq
where we assumed that near the waterfall transition the inflaton potential can be approximated by a quadratic term. 
The back-reaction term from the waterfall field in the mean field approximation is $m_0^2({\cal N}\langle \phi_i^2\rangle/\psi_0^2)\psi$, which can be safely neglected around the time of the transition for 
$\psi_0^2\gg \langle \phi_{\rm init}^2 \rangle \sim H^2$. 
The equations of motion of the waterfall fields are
\beq
\label{eq:phi_i}
\ddot\phi_i + 3 H \dot \phi_i -e^{-2Ht}\nabla^2 \phi_i + m_\phi^2(t) \phi_i=0 \, ,
\eeq
where
$
m_\phi^2(t) = -m_0^2 \left (
1- {\psi^2(t)/ \psi_c^2}
\right )
$.
We expand each waterfall field $\phi_i$ in mode-functions 
\beq
\phi_i(\vec x, t) = \int {d^3k\over (2\pi)^3} \left [
c_{k} e^{i \vec k \cdot \vec x} u_k(t) + c_{k}^\dagger e^{-i \vec k \cdot \vec x} u^*_k(t) 
\right ]
\eeq
where the mode-functions 
follow Eq.~\eqref{eq:phi_i} with $\phi_i\to u_k$ and $\nabla^2 \to -k^2$; and are initialized in the Bunch-Davies vacuum
$
u_k(t) \to {e^{-i k t/a}/( a\sqrt{2k})}
$.
As a characteristic (classical) value of the waterfall field, we can take the Root-Mean-Square (rms) value 
\beq
\phi_{i,{\rm rms}}^2(t) =\int  {dk \over 2\pi^2} k^2 \left |u_k(t) \right |^2 
\eeq
We must note that this is the rms value of  each waterfall field $\phi_i$. The total rms value of the waterfall ${\cal N}$-tuple is 
$
\phi_{\rm rms}^2(t) ={\cal N} \phi_{i,{\rm rms}}^2(t)
$.

The inflaton field, close to the waterfall transition, evolves as
$
\psi(t) = \psi_c e^{-p t},
$
with
\beq
{p \over H}=    {3\over 2} - \sqrt{{9\over 4} - {m_\psi^2\over H^2}}   \simeq {m_\psi^2\over 3H^2} +{\cal O}(m_\psi^4 / H^4),
\eeq
where the last approximation is valid for $m_\psi \ll H$.  
At late times, the gradient term in Eq.~\eqref{eq:phi_i} is exponentially suppressed, consequently all modes of the waterfall field evolve identically, leading to
\beq
\phi_i(\vec x,t) = \phi_i(\vec x,t_c) \exp \left (
\int_{t_c}^t dt' \lambda(t')
\right )
\, .
\eeq
The spatial dependence of the waterfall field is generated close to the waterfall transition (around $\psi\simeq \psi_c$) and is encoded in $\phi_i(\vec x, t_c)$, where $t_c$ is a late enough time, such that the gradient terms can be dropped for $t>t_c$.
The growth rate $\lambda(t)$ is defined through
\beq
{ \lambda(t) \over H }=  
-{3\over 2} + \sqrt{{9\over 4} + {m_0^2\over H^2}\left (1-e^{-2pt} \right)} 
\, ,
\label{eq:lambda}
\eeq
which becomes constant if $m_\psi \ll H \ll m_0$ (see Ref.~\cite{Guth:2012we} for details on the dynamics of the waterfall transition). 

The  smallest wavenumbers start growing first (around the time of the transition).
 The largest wavenumbers start growing
 when $k^2\lesssim m_0^2 a^2$, thus
 approximately $\log(k/m_0)$ $e$-folds later. 
 Modes with $k\gg m_0>H$  are thus exponentially suppressed with respect to $k\sim H$ and  do not affect the end-result~\cite{Guth:2012we} (this is further explained in the Supplemental Material). 
By numerically computing the rms value of the waterfall field for a wide range of cases, we find it to be in excellent agreement with the simple estimate
\beq
\phi_{i,{\rm rms}}^2(t) \simeq {H^2\over 2\pi^2} e^{2 \int_0^t \lambda(t') dt'} 
\, ,
\label{eq:phirmsapprox}
\eeq
which makes use of the late-time  growth rate $\lambda(t)$. We use  Eq.~\eqref{eq:phirmsapprox} to compute the  background value of the waterfall field, which induces the  growth of gauge fields through the Chern-Simons coupling.

\medskip
\noindent\textbf{Gauge field Dynamics.}
We restrict the scope of our analysis to the linear evolution equation of the gauge field $A_\mu$ and choose the Coulomb ($\partial_iA_i=0$) or temporal gauge ($A_0=0$), which are equivalent at linear order. The resulting equations for the gauge field mode-functions of the two circular polarizations ($\pm$) are 
\beq
\label{eq:Akeom}
\ddot A^\pm_k + H \dot A^\pm_k + \left ( {k^2\over a^2} \pm \sum_i {1\over \Lambda_i} {k\over a} \dot \phi_{i,{\rm rms}}  \right ) A^\pm_k =0 \, ,
\eeq
where $k$ is the comoving wavenumber. 
For Eq.~\eqref{eq:action} to be valid as a consistent Effective Field Theory (EFT), we require
 $
\Lambda_i^2 \gtrsim {\dot\phi_i} \sim H  M_{\rm Pl}$ (see e.g. Ref.~\cite{Wang:2020ioa}), 
 where $\dot\phi_i = \lambda(t)\phi_i$ at late times and
$\phi_i\le \phi_{\rm end} \sim M_{\rm Pl}$. This condition on $\Lambda_i$ is easily satisfied for the parameters of our interest (Table.~\ref{tab1}).

For $k/a< \sum_i  | \dot\phi_{\rm rms}|/\Lambda_i$, one polarization acquires a negative effective frequency-squared, leading to a tachyonic amplification \cite{Adshead:2015pva, Barnaby:2011vw, Barnaby:2011qe, Armendariz-Picon:2007gbe, Dufaux:2006ee, Garretson:1992vt}. 
Due to the exponential sensitivity of tachyonic amplification, we can replace $\sum_i \dot \phi_{i,{\rm rms}}/\Lambda_i$ with the dominant contribution, corresponding to the smallest $\Lambda_i$, $\Lambda_{i,{\rm min}}$. 
Using Eq.~\eqref{eq:phirmsapprox}, tachyonic amplification occurs for
\beq
\label{eq:kamplification}
{k\over Ha(t)} < {1\over \Lambda_{i,{\rm min}} }{1\over \sqrt{2\pi}} \lambda(t) e^{\int_0^t \lambda(t') dt'}.
\eeq
The maximum wavenumber that gets amplified, computed at the end of the waterfall phase, is
$
{k_{\rm max}/ (aH)} \simeq (\Lambda_{i,{\rm min}}H)^{-1} \lambda \dot\phi_{\rm end}
$
where 
$
\dot \phi_{\rm end} = \lambda \phi_{\rm end} \sim \lambda M_{\rm Pl}
$. 
We performed an array of numerical calculations, which consistently led to
 the peak of the gauge field spectrum being about one order of magnitude below $k_{\rm max}$ as defined earlier
 \beq
{k_{\rm peak} \over aH} \sim {\lambda\over 10H} {H\over \Lambda_{i,{\rm min}}} {M_{\rm Pl}\over H} \, .
\label{eq:kpeak}
\eeq 
As we will see (Eq.~\ref{eq:Omega}), Eq.~\eqref{eq:kpeak} allows for a  quick estimate of the peak frequency of the GWs sourced by $A_\mu$.

To optimize the GW detection prospect, we consider complete preheating into gauge fields for our benchmarks, which can be realized when the energy stored in $A_\mu$ fiels equals that of the total cosmic background energy (in the linear approximation, which we are using), i.e., $\rho_A=3H^2M_{\rm Pl}^2$. In general $\rho_A$ is computed by integrating over all wavenumbers. However, for a highly peaked spectrum we can estimate it (dropping ${\cal O}(1)$ factors) as $\rho_A\sim {\cal A}^2 k_{\rm peak}^4 /a^4 $, where ${\cal A}$ is the amplification factor. Complete preheating thus requires ${\cal A} \sim (M_{\rm Pl}/ H) \left ( {k_{\rm peak}/ aH}\right )^2$, where ${k_{\rm peak}/ aH}\sim 100$ for our chosen examples.

We can estimate ${\cal A}$ for the mode $k$ using the WKB approximation \cite{Adshead:2015pva}
as
${\cal A} = 
\exp \left [
\int_{t^-}^t dt'
\left(
-{k^2\over a^2} +{H^2\over 4}\mp {1\over \Lambda_{i,{\rm min}}}{k\over a} \dot \phi_{i,{\rm rms}}
\right ) 
\right ]$, where the integrand becomes real at $t=t^-$.
This alone does not give a simple analytical solution, but as $\dot\phi$ grows exponentially, the majority of the amplification occurs near the end of the waterfall phase, when $|\dot\phi_{\rm rms}|\sim \lambda M_{\rm Pl}$. We can then estimate $\cal A$ during the last $e$-fold as $\exp\left (
{\lambda M_{\rm Pl} \over H {\Lambda_{i,{\rm min}}}\sqrt{10}}
\right )$ (use Eq.~\ref{eq:kpeak}). With $\lambda/H ={\cal O}(1)$ and $M_{\rm Pl}/\Lambda_{i,{\rm min}}\sim 100$ (as in our benchmark examples), this leads to a sufficient amplification, as required for complete preheating.
 As in  single-field axion-inflation preheating, a reduction in the velocity of the scalar field and the $H$ can be offset by an increase in the coupling $1/\Lambda_i$, in order to have complete preheating \cite{Adshead:2015pva, Adshead:2019lbr}.

\medskip
\noindent\textbf{Gravitational Wave Signals.}
To estimate the peak GW signal, we apply the results derived in ~\cite{Giblin:2014gra} which generically apply to GWs produced by a source which peaks at a wavenumber $k_*$:

\begin{eqnarray}
\nu_{GW}^{\rm peak} &=& 2.7\times 10^{10} {{k_{*}}\over \sqrt{M_{\rm Pl} \, H} }\, {\rm Hz} \, ,
\label{eq:freq}
\\
\Omega_{\rm GW}^{\rm peak} &=& 2.3\times 10^{-4} \, \alpha^2\, \beta\, w 
\left ( {k_*\over \sigma} \right ) 
\left ({H_* \over k_*}\right )^2.
\label{eq:Omega}
\end{eqnarray}
$\alpha$ is the fraction of the energy in the GW source ($A_\mu$ in our case) relative to the Universe's total energy density and  $\beta$ encodes the anisotropy of the source. 
The equation of state of the Universe at the time of GW emission is parametrized by $w$, which is close to $1/3$ for efficient preheating into relativistic gauge fields.
The peak wavenumber of the source denoted as $k_*$ should be identified as $k_{\rm peak}/a$ as shown earlier for our case, and is proportional to the peak frequency of the produced GW spectrum. Finally, $\sigma$ is the width of the peak when the source is modeled as Gaussian (in our case $\sigma\simeq k_*$, due to the highly peaked spectrum), and 
 $H_*$ is the Hubble scale at the time of GW production, which in our case is the end of inflation. 
We focus on GW production by gauge fields, which dominate the  energy density by the end of preheating.

For numerical results, we first illustrate an example with the inflationary Hubble scale  $H={\rm MeV} = 10^{-20} M_{\rm Pl}$
and the masses of the timer and waterfall fields: $m_0=6\,H$ and $m_\phi=0.5\,H$.
 By computing  $\phi_{\rm rms}$ and considering that $\phi_{\rm init} \sim H$ and $\phi_{\rm end} \sim M_{\rm Pl}$, we find $\phi_{\rm init} / \phi_{\rm end} \sim 10^{-20}$ and $N_{\rm waterfall} = 14.2$. We choose  $\Lambda_{i, min}\simeq 10^{18}H =10^{-2}M_{\rm Pl} $ in order to allow complete preheating into gauge fields by the end of the waterfall phase, while  avoiding significant back-reaction. 
Following Eqs.~\eqref{eq:freq} and \eqref{eq:Omega}, the peak frequency of the GW's is $\nu_{\rm {GW}}^{\rm peak}\simeq 100~{\rm Hz}$ and the peak amplitude is estimated as
$\Omega_{\rm GW}^{\rm peak} \simeq 10^{-8}\alpha^2 \beta $, where we set $w=1/3$ which is consistent with efficient preheating to relativistic $A_\mu$ fields. The precise values of $\alpha$ and $\beta$ require computation using lattice simulations. However, we can make reasonable estimates. $\alpha$ is expected to be close to unity, due to the efficient tachyonic preheating leading to a universe dominated by $A_\mu$. The anisotropy encoded in $\beta$ is harder to quantify. However, the result from lattice simulations of $U(1)$ gauge field production in high-scale single-field axion inflation should give a close approximation, which shows $\beta={\cal O}(0.1)$ \cite{Adshead:2015pva}. Putting all together, we find and $\Omega_{\rm GW}\simeq 10^{-10}$. 
$\nu_{\rm {GW}}^{\rm peak}$ in this example is within the LIGO band, but ET or CE may be needed to reach sufficient sensitivity to $\Omega_{\rm GW}$. The projected sensitivity of LIGO A+ is slightly above, yet within an order of magnitude of our estimated signal, hence more detailed analysis is required to reach a definite conclusion.

\begin{figure}
\includegraphics[width=\columnwidth]{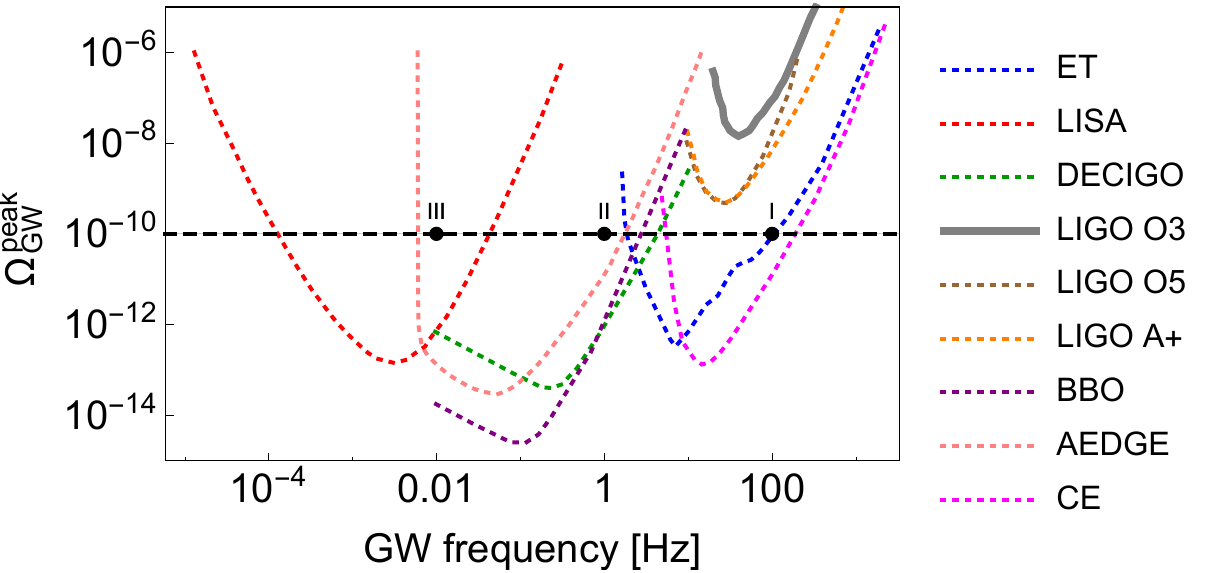}
\caption{The estimated peak amplitudes (shown as solid black dots) of GW signal $\Omega_{\rm GW}^{\rm peak}$ for benchmark parameter choices as given in Table~\ref{tab1}. The grey solid curve shows the current exclusion by LIGO. Dashed curves show the sensitivity forecasts for a variety of future GW experiments.  
}
\label{fig:experiments}
\end{figure}

Table~\ref{tab1} and Fig.~\ref{fig:experiments} illustrate the above example and several other benchmarks that are within sensitivity of various upcoming GW experiments.
An additional example within reach of pulsar timing experiments such as SKA \cite{Janssen:2014dka} can be given with: $H=10^{-40}\, M_{\rm Pl}$, $m_\psi=H/2$ and $m_0= 15H$, leading to $N_{\rm wf}=10$. Further choosing $\Lambda=10^{37}H$ 
 leads to $\nu_{\rm GW}^{\rm peak}=10^{-8}\, {\rm Hz}$ and $\Omega_{\rm GW}^{\rm peak}\sim 10^{-11}$. However, we did not include this example in Table~\ref{tab1} and Fig.~\ref{fig:experiments}, because of the huge mass of the produced BH's ($\sim 10^{22}\, M_\odot$), which is subject to severe constraint by CMB data \cite{Poulin:2017bwe} (see next section for details on BH formation).

 While $\Omega_{\rm GW}^{\rm peak}\sim 10^{-10}$ is a ballpark for our benchmark examples, it vary depending on several parameters.
It can be reduced, if preheating to gauge fields is incomplete, thus placing only a fraction of the energy density of the universe in $A_\mu$'s which source GW's, and consequently reducing the value of $\alpha$ in Eq.~\eqref{eq:Omega}. The peak frequency remains largely unchanged by this, as long as $H$ and the shape of the $A_\mu$ spectrum remain qualitatively similar. 
Increasing $\Omega_{\rm GW}^{\rm peak}$ is more difficult. Since we have chosen optimistic values for the parameters $\alpha,\beta,w$ in Eq.~\eqref{eq:Omega}, one would need to reduce the ratio $k_*/H$. In all our examples, $k_*/H\sim 100$, as required for complete preheating. A thorough parameter scan and detailed simulations can provide a more accurate value of $k_*/H$, possibly leading to a mild increase of $\Omega_{\rm GW}^{\rm peak}$.
Furthermore, the $A_\mu$ spectrum shape may deviate from the form we computed through rescattering processes, which can be captured only through lattice simulations \cite{Adshead:2016iae}.

Finally, the GWs inherits the helical polarization of their source field, $A_\mu$, which originates from the axial coupling of $\phi_i$ to the CS term involving $A_\mu$ \cite{Okano:2020uyr}.

\medskip
\noindent\textbf{Formation of Primordial Black Hole (PBH).} 
Hybrid inflation in general may lead to a large spike in the density perturbation spectrum which results in PBH formation. It is important to check if our benchmarks yielding observable GW signals are compatible with the current constraints from PBH searches \cite{Katz:2018zrn,Smyth:2019whb,Takhistov:2020ssb,Villanueva-Domingo:2021spv} with potential complementarity. There are different approaches for computing such a spectrum arising from a waterfall transition (e.g.~\cite{Randall:1995dj, Guth:2012we, Halpern:2014mca, Lyth:2010zq, Lyth:2012yp, Clesse:2015wea}). 
The ambiguities arise from the purely quantum nature of the waterfall field destabilization, which  leads to the absence of a well-defined classical trajectory~\cite{Guth:2012we, Halpern:2014mca, Randall:1995dj}.
For a  simple intuitive approximation, one can 
identify the analogue of a classical trajectory with the rms  value of the $\phi$ fluctuations,  
which is given by
$
\delta\phi(k) = 
 {k\over 2\pi}  |u_k|
$;
then compute the curvature perturbation, through the usual single-field relation $\zeta \approx {\delta\phi(k) / \dot\phi_{\rm rms}}$.
While this simple approximation may underestimate $\zeta$ by an ${\cal O}(1)$ factor (see \cite{Guth:2012we, Lyth:2010zq, Lyth:2012yp} for details), it provides intuitive insights into the problem.

The mass of a PBH whose formation is triggered by a
density fluctuation exiting the Hubble radius $N_*$ e-folds before the end of inflation, is given by
$
M =(M_{\rm Pl}^2 / H_*)e^{2N_*}
$.
Approximating the probability distribution of density
perturbations as Gaussian, 
 the fraction
 of the Universe's energy collapsing into PBH's of mass $M$ at the time of formation $\beta_{\rm BH}$ is 
\beq
\label{eq:betaBH}
\beta_{\rm BH}(M) = \erfc \left (
{\zeta_c \over \sqrt{2}\sigma}
\right )
\simeq {{\sqrt 2}\sigma\over \sqrt{\pi} \zeta_c}e^{-\zeta_c^2 /(2\sigma^2)},
\eeq
where the last equality  holds for $\sigma \ll \zeta_c$ and $\zeta_c$ is the critical overdensity required for black hole formation. The variance $\sigma$ is related to the scalar power spectrum as $\sigma^2 = \langle \zeta^2\rangle = {\cal P}_{\zeta}$ for each wavenumber.
The mass of the BH relates to the Hubble scale of the universe when a certain mode $k$ re-enters horizon and triggers BH formation, and thus $\sigma(k)$ determines $\beta_{\rm BH}(M)$.
The exact value of  $\zeta_c$ must be computed numerically and depends on the exact shape of the over-density~\cite{Musco:2004ak, Niemeyer:1999ak, Shibata:1999zs, Yoo:2018kvb, Yoo:2020lmg, Harada:2013epa}, while the canonical expectation lies in the range of
$0.01\lesssim \zeta_c< 1$ \cite{Harada:2013epa}.

During the radiation dominated era, the evolution equation $d\beta_{\rm BH}(M,t) /dt \simeq H(t) \beta_{\rm BH}(M,t)$ shows the redshift of the fraction of the universe that has collapsed into BH's.
The  abundance of the PBH's during matter-radiation equality is,  $\Omega_{\rm BH}(t_{\rm eq}) = \int_0^M \beta_{\rm BH}(M',t_{\rm eq}) d\log M'$. After matter-radiation equality, the evolution of the BH abundance follows that of ordinary matter.

 In Table~\ref{tab1} we show the typical PBH mass predicted with the benchmark parameters. This
corresponds to the BH's formed by fluctuations which exited the horizon at the time of the waterfall transition at $\psi=\psi_c$.
 But we do not explicitly show $\Omega_{\rm BH}$ because of the large uncertainty on $\zeta_c$ which requires dedicate simulation to clarify. Ref.~\cite{Clesse:2015wea}
showed that by taking the critical overdensity $\zeta_c$ as a free parameter, one is generally able to fit a desired value of $\Omega_{\rm BH}$ with typical hybrid inflation power spectra (in some cases requiring $\zeta_c$ to deviate by orders of magnitude from the canonical expectation). For example, using Eq.~\eqref{eq:betaBH} with $\zeta_c\sim 0.2$, $H=10^{-20}M_{\rm Pl}$, $m_0=15H$ and $m_\psi=H/2$, we find $M_{\rm BH}\sim 10^{-13} M_\odot$  and $\Omega_{\rm BH}$ can saturate the observed dark matter (DM) abundance today or be an appreciable fraction of DM while being compatible with existing bounds from PBH searches. The last example in Table~\ref{tab1} leads to BH's with masses of ${\cal O}(10) M_\odot$.  In order to satisfy  constraints from gravitational lensing, BH's can only be a fraction of DM $\Omega_{\rm BH}\lesssim 0.1~\Omega_{\rm DM}$, which requires $\zeta_c\gtrsim 0.2$. 
This example also demonstrates that such models can produce PBH's whose late-time inspirals/mergers events are detectable by experiments such as LIGO and LISA in form of transient GW signals.
The sensitive dependence of the BH characteristics on the potential parameters makes a detailed parameter scan very intriguing. The exponential sensitivity of the BH formation process on the scalar over-densities calls for detailed simulations of BH formation in hybrid inflation, taking into account the  distribution and form (ellipticity) of the large overdensities \cite{Bloomfield:2016civ, Bloomfield:2018oku}.

\begin{table}
	\begin{tabular}{|c|c|c|c|c|c|c|}
		\hline	
		$H/M_{\rm Pl}$          & $m_0$    &  $\Lambda/H$     &  $N_{\rm wf}$     &  $\nu^{\rm peak}_{\rm GW}$          & $\Omega^{\rm peak}_{\rm GW}$                        & $M_{\rm BH} $                               \\ \hline\hline
	
			$10^{-20}$          & $6\,H$      &  $10^{18}$        &  $14.2$                &  $100 \, {\rm  Hz}$          & $10^{-10}$                        & $10^{-5}M_\odot $                
				\\
			\hline
			$10^{-20}$          & $15\,H$      &  $10^{18}$  &  $6.2$      &  $100 \, {\rm  Hz}$          & $10^{-10}$                        & $10^{-13} M_\odot $                
			\\
			\hline
			$10^{-24}$       & $7\,H$      &  $10^{22}$    &  $14$    &  $1 \, {\rm  Hz}$          & $10^{-10}$                        & $0.1 M_\odot $            
			\\
			\hline
			$10^{-30}$          & $8\,H$      &  $10^{27}$  &  $14.5$      &  $10^{-3} \, {\rm  Hz}$          & $10^{-10}$                        & $10^5 M_\odot $  
		
		\\
			\hline
			$10^{-30}$          & $12\,H$      &  $10^{27}$  &  $10$      &  $10^{-3} \, {\rm  Hz}$          & $10^{-10}$                        & $10 M_\odot $  
			\\
			
	\hline

	\end{tabular}
	\caption{The parameters and observables for four characteristic choices (we fix $m_\phi=H/2$), leading to GW production within frequency coverage of LIGO, LISA and midband experiments.
	$M_{\rm BH}$ corresponds to the peak of the BH mass distribution.
	}
	\label{tab1}
\end{table}
%
%


\medskip
\noindent\textbf{Summary and  Discussion.}
In this \textit{Letter} 
we propose a version of hybrid inflation model involving an axion-like waterfall field coupling to an Abelian gauge field via a Chern-Simons term which triggers preheating. The preheating process in this model can produce  characteristic, chiral GW signatures with signal strength and frequency peak within detectable range in the near future. The helically polarized nature of the GW signal predicted from the scenario can also be probed by foreseeable GW experiments \cite{Domcke:2019zls, Smith:2016jqs, Belgacem:2020nda}, which may help distinguish itself from SGWB due to other sources such as phase transitions, cosmic strings or astrophysical background (e.g.~\cite{Grojean:2006bp, Caprini:2009yp,Vachaspati:1984gt,Auclair:2019wcv,Barish:2020vmy,Cui:2017ufi,Cui:2018rwi,Caprini:2018mtu,Caprini:2019egz}).
Other proposed mechanisms generating a helical GW signature include, e.g., those considered in \cite{Thorne:2017jft, Caldwell:2017chz, Dimastrogiovanni:2016fuu, Fujita:2018ndp,DAmico:2020euu, DAmico:2021vka}.

Our finding is an important complement to the literature which so far suggested that GW signal from preheating, albeit appealing, is generally not detectable with near future GW detectors. We discuss the comparison with earlier literature in the Supplemental Material. Furthermore, the scenario we consider leads to the production of PBH's that can be (a fraction of) dark matter and their mergers can lead to a complementary channel of GW signatures. In fact, following Table~\ref{tab1}, correlating the SGWB and late-time transient GW signals could
allow us to break degeneracy and better determine model parameters.

The Abelian gauge field $A_\mu$ in this model in principle can be the SM photon. However, in the desirable case where the GW signal frequency is low enough to be detectable, the waterfall field $\phi$ has a mass of $O(H)\lesssim1 \, {\rm GeV}$ and thus is subject to the strong constraints on axion-photon coupling \cite{ParticleDataGroup:2020ssz}. The only region in this mass range that remains inaccessible by current terrestrial probes is the ``cosmological triangle" (axion mass $\sim {\cal O}({\rm GeV})$ and $1/\Lambda\sim 10^{-5}$ GeV$^{-1}$) \cite{Cadamuro:2010cz, Cadamuro:2011fd,Depta:2020wmr}, which we found to be in conflict with the EFT condition for the $A_\mu$-$\phi$ coupling in our model. Nevertheless, a viable and interesting possibility is that $A_\mu$ is a dark photon residing in a dark matter sector, and the SM particles can be populated via interactions with the dark states. The detailed realization is beyond the scope of this work and we leave it for future investigation.

In our studies, we approximate the waterfall transition with an inverted quadratic potential, neglecting the effects near the end, where the stabilizing higher order (e.g. quartic) terms become relevant. Furthermore, we use the linear evolution equations for the gauge fields, which inevitably break down when the energy density of the gauge fields become similar to that of the inflationary sector. However, by properly choosing our parameters, we can adjust the time of the completion of preheating to coincide with the end of the waterfall period, thus pushing the back-reaction regime to the end of preheating, so that our approximation is reliable. Nevertheless, the importance of the prospect of probing inflation/(p)reheating era with GWs warrants a detailed lattice simulation based study in order to go beyond the limitations of our simplified approach and identify the spectral shape of the GW signal originated from this model. This is a non-trivial undertaking, since the simulation would need to resolve both the dynamics of the waterfall transition, as well as the transfer of energy to the gauge fields, which occur at different times. We thus defer such a detailed simulation for future work.

To conclude, we demonstrated that a simple scenario of inflationary preheating can produce observable and characteristic GW signals, allowing us to probe the very early stage of the Universe prior to the hot Big Bang. Complemented by dedicated follow-up studies, this work will lead to new well-motivated targets for GW searches with various current or upcoming experiments, such as LISA, ET, CE,  AEDGE, DECIGO, BBO, and possibly LIGO A+.

\begin{acknowledgments}
\section*{Acknowledgements}
We thank the KITP (supported in part by NSF under Grant No. NSF PHY-1748958) for hospitality.
YC is supported in part by the US Department of Energy under award number DE-SC0008541.
EIS is supported by
a fellowship from ``la Caixa'' Foundation (ID 100010434) and from the European Union's Horizon
2020 research and innovation programme under the Marie Sk\l odowska-Curie grant agreement No 
847648, the fellowship code is LCF/BQ/PI20/11760021.
\end{acknowledgments}

\bibliography{prehGW}

\providecommand{\noopsort}[1]{}\providecommand{\singleletter}[1]{#1}%
\begin{thebibliography}{87}%
\makeatletter
\providecommand \@ifxundefined [1]{%
 \@ifx{#1\undefined}
}%
\providecommand \@ifnum [1]{%
 \ifnum #1\expandafter \@firstoftwo
 \else \expandafter \@secondoftwo
 \fi
}%
\providecommand \@ifx [1]{%
 \ifx #1\expandafter \@firstoftwo
 \else \expandafter \@secondoftwo
 \fi
}%
\providecommand \natexlab [1]{#1}%
\providecommand \enquote  [1]{``#1''}%
\providecommand \bibnamefont  [1]{#1}%
\providecommand \bibfnamefont [1]{#1}%
\providecommand \citenamefont [1]{#1}%
\providecommand \href@noop [0]{\@secondoftwo}%
\providecommand \href [0]{\begingroup \@sanitize@url \@href}%
\providecommand \@href[1]{\@@startlink{#1}\@@href}%
\providecommand \@@href[1]{\endgroup#1\@@endlink}%
\providecommand \@sanitize@url [0]{\catcode `\\12\catcode `\$12\catcode
  `\&12\catcode `\#12\catcode `\^12\catcode `\_12\catcode `\%12\relax}%
\providecommand \@@startlink[1]{}%
\providecommand \@@endlink[0]{}%
\providecommand \url  [0]{\begingroup\@sanitize@url \@url }%
\providecommand \@url [1]{\endgroup\@href {#1}{\urlprefix }}%
\providecommand \urlprefix  [0]{URL }%
\providecommand \Eprint [0]{\href }%
\providecommand \doibase [0]{http://dx.doi.org/}%
\providecommand \selectlanguage [0]{\@gobble}%
\providecommand \bibinfo  [0]{\@secondoftwo}%
\providecommand \bibfield  [0]{\@secondoftwo}%
\providecommand \translation [1]{[#1]}%
\providecommand \BibitemOpen [0]{}%
\providecommand \bibitemStop [0]{}%
\providecommand \bibitemNoStop [0]{.\EOS\space}%
\providecommand \EOS [0]{\spacefactor3000\relax}%
\providecommand \BibitemShut  [1]{\csname bibitem#1\endcsname}%
\let\auto@bib@innerbib\@empty
\bibitem [{\citenamefont {Guth}\ and\ \citenamefont
  {Kaiser}(2005)}]{Guth:2005zr}%
  \BibitemOpen
  \bibfield  {author} {\bibinfo {author} {\bibfnamefont {A.~H.}\ \bibnamefont
  {Guth}}\ and\ \bibinfo {author} {\bibfnamefont {D.~I.}\ \bibnamefont
  {Kaiser}},\ }\href {\doibase 10.1126/science.1107483} {\bibfield  {journal}
  {\bibinfo  {journal} {Science}\ }\textbf {\bibinfo {volume} {307}},\ \bibinfo
  {pages} {884} (\bibinfo {year} {2005})},\ \Eprint
  {http://arxiv.org/abs/astro-ph/0502328} {arXiv:astro-ph/0502328} \BibitemShut
  {NoStop}%
\bibitem [{\citenamefont {Linde}(1983)}]{Linde:1983gd}%
  \BibitemOpen
  \bibfield  {author} {\bibinfo {author} {\bibfnamefont {A.~D.}\ \bibnamefont
  {Linde}},\ }\href {\doibase 10.1016/0370-2693(83)90837-7} {\bibfield
  {journal} {\bibinfo  {journal} {Phys. Lett. B}\ }\textbf {\bibinfo {volume}
  {129}},\ \bibinfo {pages} {177} (\bibinfo {year} {1983})}\BibitemShut
  {NoStop}%
\bibitem [{\citenamefont {Akrami}\ \emph {et~al.}(2020)\citenamefont {Akrami}
  \emph {et~al.}}]{Planck:2018jri}%
  \BibitemOpen
  \bibfield  {author} {\bibinfo {author} {\bibfnamefont {Y.}~\bibnamefont
  {Akrami}} \emph {et~al.} (\bibinfo {collaboration} {Planck}),\ }\href
  {\doibase 10.1051/0004-6361/201833887} {\bibfield  {journal} {\bibinfo
  {journal} {Astron. Astrophys.}\ }\textbf {\bibinfo {volume} {641}},\ \bibinfo
  {pages} {A10} (\bibinfo {year} {2020})},\ \Eprint
  {http://arxiv.org/abs/1807.06211} {arXiv:1807.06211 [astro-ph.CO]}
  \BibitemShut {NoStop}%
\bibitem [{\citenamefont {Albrecht}\ \emph {et~al.}(1982)\citenamefont
  {Albrecht}, \citenamefont {Steinhardt}, \citenamefont {Turner},\ and\
  \citenamefont {Wilczek}}]{Albrecht:1982mp}%
  \BibitemOpen
  \bibfield  {author} {\bibinfo {author} {\bibfnamefont {A.}~\bibnamefont
  {Albrecht}}, \bibinfo {author} {\bibfnamefont {P.~J.}\ \bibnamefont
  {Steinhardt}}, \bibinfo {author} {\bibfnamefont {M.~S.}\ \bibnamefont
  {Turner}}, \ and\ \bibinfo {author} {\bibfnamefont {F.}~\bibnamefont
  {Wilczek}},\ }\href {\doibase 10.1103/PhysRevLett.48.1437} {\bibfield
  {journal} {\bibinfo  {journal} {Phys. Rev. Lett.}\ }\textbf {\bibinfo
  {volume} {48}},\ \bibinfo {pages} {1437} (\bibinfo {year}
  {1982})}\BibitemShut {NoStop}%
\bibitem [{\citenamefont {Kofman}\ \emph {et~al.}(1997)\citenamefont {Kofman},
  \citenamefont {Linde},\ and\ \citenamefont {Starobinsky}}]{Kofman:1997yn}%
  \BibitemOpen
  \bibfield  {author} {\bibinfo {author} {\bibfnamefont {L.}~\bibnamefont
  {Kofman}}, \bibinfo {author} {\bibfnamefont {A.~D.}\ \bibnamefont {Linde}}, \
  and\ \bibinfo {author} {\bibfnamefont {A.~A.}\ \bibnamefont {Starobinsky}},\
  }\href {\doibase 10.1103/PhysRevD.56.3258} {\bibfield  {journal} {\bibinfo
  {journal} {Phys. Rev. D}\ }\textbf {\bibinfo {volume} {56}},\ \bibinfo
  {pages} {3258} (\bibinfo {year} {1997})},\ \Eprint
  {http://arxiv.org/abs/hep-ph/9704452} {arXiv:hep-ph/9704452} \BibitemShut
  {NoStop}%
\bibitem [{\citenamefont {Kofman}\ \emph {et~al.}(1994)\citenamefont {Kofman},
  \citenamefont {Linde},\ and\ \citenamefont {Starobinsky}}]{Kofman:1994rk}%
  \BibitemOpen
  \bibfield  {author} {\bibinfo {author} {\bibfnamefont {L.}~\bibnamefont
  {Kofman}}, \bibinfo {author} {\bibfnamefont {A.~D.}\ \bibnamefont {Linde}}, \
  and\ \bibinfo {author} {\bibfnamefont {A.~A.}\ \bibnamefont {Starobinsky}},\
  }\href {\doibase 10.1103/PhysRevLett.73.3195} {\bibfield  {journal} {\bibinfo
   {journal} {Phys. Rev. Lett.}\ }\textbf {\bibinfo {volume} {73}},\ \bibinfo
  {pages} {3195} (\bibinfo {year} {1994})},\ \Eprint
  {http://arxiv.org/abs/hep-th/9405187} {arXiv:hep-th/9405187} \BibitemShut
  {NoStop}%
\bibitem [{\citenamefont {Amin}\ \emph {et~al.}(2014)\citenamefont {Amin},
  \citenamefont {Hertzberg}, \citenamefont {Kaiser},\ and\ \citenamefont
  {Karouby}}]{Amin:2014eta}%
  \BibitemOpen
  \bibfield  {author} {\bibinfo {author} {\bibfnamefont {M.~A.}\ \bibnamefont
  {Amin}}, \bibinfo {author} {\bibfnamefont {M.~P.}\ \bibnamefont {Hertzberg}},
  \bibinfo {author} {\bibfnamefont {D.~I.}\ \bibnamefont {Kaiser}}, \ and\
  \bibinfo {author} {\bibfnamefont {J.}~\bibnamefont {Karouby}},\ }\href
  {\doibase 10.1142/S0218271815300037} {\bibfield  {journal} {\bibinfo
  {journal} {Int. J. Mod. Phys. D}\ }\textbf {\bibinfo {volume} {24}},\
  \bibinfo {pages} {1530003} (\bibinfo {year} {2014})},\ \Eprint
  {http://arxiv.org/abs/1410.3808} {arXiv:1410.3808 [hep-ph]} \BibitemShut
  {NoStop}%
\bibitem [{\citenamefont {Adshead}\ \emph
  {et~al.}(2020{\natexlab{a}})\citenamefont {Adshead}, \citenamefont {Giblin},
  \citenamefont {Pieroni},\ and\ \citenamefont {Weiner}}]{Adshead:2019lbr}%
  \BibitemOpen
  \bibfield  {author} {\bibinfo {author} {\bibfnamefont {P.}~\bibnamefont
  {Adshead}}, \bibinfo {author} {\bibfnamefont {J.~T.}\ \bibnamefont {Giblin}},
  \bibinfo {author} {\bibfnamefont {M.}~\bibnamefont {Pieroni}}, \ and\
  \bibinfo {author} {\bibfnamefont {Z.~J.}\ \bibnamefont {Weiner}},\ }\href
  {\doibase 10.1103/PhysRevD.101.083534} {\bibfield  {journal} {\bibinfo
  {journal} {Phys. Rev. D}\ }\textbf {\bibinfo {volume} {101}},\ \bibinfo
  {pages} {083534} (\bibinfo {year} {2020}{\natexlab{a}})},\ \Eprint
  {http://arxiv.org/abs/1909.12842} {arXiv:1909.12842 [astro-ph.CO]}
  \BibitemShut {NoStop}%
\bibitem [{\citenamefont {Adshead}\ \emph {et~al.}(2018)\citenamefont
  {Adshead}, \citenamefont {Giblin},\ and\ \citenamefont
  {Weiner}}]{Adshead:2018doq}%
  \BibitemOpen
  \bibfield  {author} {\bibinfo {author} {\bibfnamefont {P.}~\bibnamefont
  {Adshead}}, \bibinfo {author} {\bibfnamefont {J.~T.}\ \bibnamefont {Giblin}},
  \ and\ \bibinfo {author} {\bibfnamefont {Z.~J.}\ \bibnamefont {Weiner}},\
  }\href {\doibase 10.1103/PhysRevD.98.043525} {\bibfield  {journal} {\bibinfo
  {journal} {Phys. Rev. D}\ }\textbf {\bibinfo {volume} {98}},\ \bibinfo
  {pages} {043525} (\bibinfo {year} {2018})},\ \Eprint
  {http://arxiv.org/abs/1805.04550} {arXiv:1805.04550 [astro-ph.CO]}
  \BibitemShut {NoStop}%
\bibitem [{\citenamefont {Lozanov}\ and\ \citenamefont
  {Amin}(2019)}]{Lozanov:2019ylm}%
  \BibitemOpen
  \bibfield  {author} {\bibinfo {author} {\bibfnamefont {K.~D.}\ \bibnamefont
  {Lozanov}}\ and\ \bibinfo {author} {\bibfnamefont {M.~A.}\ \bibnamefont
  {Amin}},\ }\href {\doibase 10.1103/PhysRevD.99.123504} {\bibfield  {journal}
  {\bibinfo  {journal} {Phys. Rev. D}\ }\textbf {\bibinfo {volume} {99}},\
  \bibinfo {pages} {123504} (\bibinfo {year} {2019})},\ \Eprint
  {http://arxiv.org/abs/1902.06736} {arXiv:1902.06736 [astro-ph.CO]}
  \BibitemShut {NoStop}%
\bibitem [{\citenamefont {Antusch}\ \emph {et~al.}(2017)\citenamefont
  {Antusch}, \citenamefont {Cefala},\ and\ \citenamefont
  {Orani}}]{Antusch:2016con}%
  \BibitemOpen
  \bibfield  {author} {\bibinfo {author} {\bibfnamefont {S.}~\bibnamefont
  {Antusch}}, \bibinfo {author} {\bibfnamefont {F.}~\bibnamefont {Cefala}}, \
  and\ \bibinfo {author} {\bibfnamefont {S.}~\bibnamefont {Orani}},\ }\href
  {\doibase 10.1103/PhysRevLett.118.011303} {\bibfield  {journal} {\bibinfo
  {journal} {Phys. Rev. Lett.}\ }\textbf {\bibinfo {volume} {118}},\ \bibinfo
  {pages} {011303} (\bibinfo {year} {2017})},\ \bibinfo {note} {[Erratum:
  Phys.Rev.Lett. 120, 219901 (2018)]},\ \Eprint
  {http://arxiv.org/abs/1607.01314} {arXiv:1607.01314 [astro-ph.CO]}
  \BibitemShut {NoStop}%
\bibitem [{\citenamefont {Amin}\ \emph {et~al.}(2018)\citenamefont {Amin},
  \citenamefont {Braden}, \citenamefont {Copeland}, \citenamefont {Giblin},
  \citenamefont {Solorio}, \citenamefont {Weiner},\ and\ \citenamefont
  {Zhou}}]{Amin:2018xfe}%
  \BibitemOpen
  \bibfield  {author} {\bibinfo {author} {\bibfnamefont {M.~A.}\ \bibnamefont
  {Amin}}, \bibinfo {author} {\bibfnamefont {J.}~\bibnamefont {Braden}},
  \bibinfo {author} {\bibfnamefont {E.~J.}\ \bibnamefont {Copeland}}, \bibinfo
  {author} {\bibfnamefont {J.~T.}\ \bibnamefont {Giblin}}, \bibinfo {author}
  {\bibfnamefont {C.}~\bibnamefont {Solorio}}, \bibinfo {author} {\bibfnamefont
  {Z.~J.}\ \bibnamefont {Weiner}}, \ and\ \bibinfo {author} {\bibfnamefont
  {S.-Y.}\ \bibnamefont {Zhou}},\ }\href {\doibase 10.1103/PhysRevD.98.024040}
  {\bibfield  {journal} {\bibinfo  {journal} {Phys. Rev. D}\ }\textbf {\bibinfo
  {volume} {98}},\ \bibinfo {pages} {024040} (\bibinfo {year} {2018})},\
  \Eprint {http://arxiv.org/abs/1803.08047} {arXiv:1803.08047 [astro-ph.CO]}
  \BibitemShut {NoStop}%
\bibitem [{\citenamefont {Hiramatsu}\ \emph {et~al.}(2021)\citenamefont
  {Hiramatsu}, \citenamefont {Sfakianakis},\ and\ \citenamefont
  {Yamaguchi}}]{Hiramatsu:2020obh}%
  \BibitemOpen
  \bibfield  {author} {\bibinfo {author} {\bibfnamefont {T.}~\bibnamefont
  {Hiramatsu}}, \bibinfo {author} {\bibfnamefont {E.~I.}\ \bibnamefont
  {Sfakianakis}}, \ and\ \bibinfo {author} {\bibfnamefont {M.}~\bibnamefont
  {Yamaguchi}},\ }\href {\doibase 10.1007/JHEP03(2021)021} {\bibfield
  {journal} {\bibinfo  {journal} {JHEP}\ }\textbf {\bibinfo {volume} {03}},\
  \bibinfo {pages} {021} (\bibinfo {year} {2021})},\ \Eprint
  {http://arxiv.org/abs/2011.12201} {arXiv:2011.12201 [hep-ph]} \BibitemShut
  {NoStop}%
\bibitem [{\citenamefont {Garcia-Bellido}\ and\ \citenamefont
  {Figueroa}(2007)}]{Garcia-Bellido:2007nns}%
  \BibitemOpen
  \bibfield  {author} {\bibinfo {author} {\bibfnamefont {J.}~\bibnamefont
  {Garcia-Bellido}}\ and\ \bibinfo {author} {\bibfnamefont {D.~G.}\
  \bibnamefont {Figueroa}},\ }\href {\doibase 10.1103/PhysRevLett.98.061302}
  {\bibfield  {journal} {\bibinfo  {journal} {Phys. Rev. Lett.}\ }\textbf
  {\bibinfo {volume} {98}},\ \bibinfo {pages} {061302} (\bibinfo {year}
  {2007})},\ \Eprint {http://arxiv.org/abs/astro-ph/0701014}
  {arXiv:astro-ph/0701014} \BibitemShut {NoStop}%
\bibitem [{\citenamefont {Garcia-Bellido}\ \emph {et~al.}(2008)\citenamefont
  {Garcia-Bellido}, \citenamefont {Figueroa},\ and\ \citenamefont
  {Sastre}}]{Garcia-Bellido:2007fiu}%
  \BibitemOpen
  \bibfield  {author} {\bibinfo {author} {\bibfnamefont {J.}~\bibnamefont
  {Garcia-Bellido}}, \bibinfo {author} {\bibfnamefont {D.~G.}\ \bibnamefont
  {Figueroa}}, \ and\ \bibinfo {author} {\bibfnamefont {A.}~\bibnamefont
  {Sastre}},\ }\href {\doibase 10.1103/PhysRevD.77.043517} {\bibfield
  {journal} {\bibinfo  {journal} {Phys. Rev. D}\ }\textbf {\bibinfo {volume}
  {77}},\ \bibinfo {pages} {043517} (\bibinfo {year} {2008})},\ \Eprint
  {http://arxiv.org/abs/0707.0839} {arXiv:0707.0839 [hep-ph]} \BibitemShut
  {NoStop}%
\bibitem [{\citenamefont {Dufaux}\ \emph {et~al.}(2010)\citenamefont {Dufaux},
  \citenamefont {Figueroa},\ and\ \citenamefont
  {Garcia-Bellido}}]{Dufaux:2010cf}%
  \BibitemOpen
  \bibfield  {author} {\bibinfo {author} {\bibfnamefont {J.-F.}\ \bibnamefont
  {Dufaux}}, \bibinfo {author} {\bibfnamefont {D.~G.}\ \bibnamefont
  {Figueroa}}, \ and\ \bibinfo {author} {\bibfnamefont {J.}~\bibnamefont
  {Garcia-Bellido}},\ }\href {\doibase 10.1103/PhysRevD.82.083518} {\bibfield
  {journal} {\bibinfo  {journal} {Phys. Rev. D}\ }\textbf {\bibinfo {volume}
  {82}},\ \bibinfo {pages} {083518} (\bibinfo {year} {2010})},\ \Eprint
  {http://arxiv.org/abs/1006.0217} {arXiv:1006.0217 [astro-ph.CO]} \BibitemShut
  {NoStop}%
\bibitem [{Note1()}]{Note1}%
  \BibitemOpen
  \bibinfo {note} {The scenario in \cite {Garcia-Bellido:2007nns,
  Garcia-Bellido:2007fiu} can lead to GW signal of $\Omega _{\protect \rm
  GW}\sim 10^{-15}$ within the frequency range of the relatively futuristic BBO
  experiment, yet much weaker than the typical prediction in our model, while
  the one in \cite {Dufaux:2010cf} can potentially lead to a detectable
  secondary peak in the advanced LIGO band.}\BibitemShut {Stop}%
\bibitem [{\citenamefont {Binetruy}\ \emph {et~al.}(2012)\citenamefont
  {Binetruy}, \citenamefont {Bohe}, \citenamefont {Caprini},\ and\
  \citenamefont {Dufaux}}]{Binetruy:2012ze}%
  \BibitemOpen
  \bibfield  {author} {\bibinfo {author} {\bibfnamefont {P.}~\bibnamefont
  {Binetruy}}, \bibinfo {author} {\bibfnamefont {A.}~\bibnamefont {Bohe}},
  \bibinfo {author} {\bibfnamefont {C.}~\bibnamefont {Caprini}}, \ and\
  \bibinfo {author} {\bibfnamefont {J.-F.}\ \bibnamefont {Dufaux}},\ }\href
  {\doibase 10.1088/1475-7516/2012/06/027} {\bibfield  {journal} {\bibinfo
  {journal} {JCAP}\ }\textbf {\bibinfo {volume} {1206}},\ \bibinfo {pages}
  {027} (\bibinfo {year} {2012})},\ \Eprint {http://arxiv.org/abs/1201.0983}
  {arXiv:1201.0983 [gr-qc]} \BibitemShut {NoStop}%
\bibitem [{\citenamefont {Maggiore}\ \emph {et~al.}(2020)\citenamefont
  {Maggiore} \emph {et~al.}}]{EinsteinTelescope}%
  \BibitemOpen
  \bibfield  {author} {\bibinfo {author} {\bibfnamefont {M.}~\bibnamefont
  {Maggiore}} \emph {et~al.},\ }\href {\doibase 10.1088/1475-7516/2020/03/050}
  {\bibfield  {journal} {\bibinfo  {journal} {JCAP}\ }\textbf {\bibinfo
  {volume} {03}},\ \bibinfo {pages} {050} (\bibinfo {year} {2020})},\ \Eprint
  {http://arxiv.org/abs/1912.02622} {arXiv:1912.02622 [astro-ph.CO]}
  \BibitemShut {NoStop}%
\bibitem [{\citenamefont {Reitze}\ \emph {et~al.}(2019)\citenamefont {Reitze}
  \emph {et~al.}}]{CosmicExplorer}%
  \BibitemOpen
  \bibfield  {author} {\bibinfo {author} {\bibfnamefont {D.}~\bibnamefont
  {Reitze}} \emph {et~al.},\ }\href@noop {} {\bibfield  {journal} {\bibinfo
  {journal} {Bull. Am. Astron. Soc.}\ }\textbf {\bibinfo {volume} {51}},\
  \bibinfo {pages} {035} (\bibinfo {year} {2019})},\ \Eprint
  {http://arxiv.org/abs/1907.04833} {arXiv:1907.04833 [astro-ph.IM]}
  \BibitemShut {NoStop}%
\bibitem [{\citenamefont {El-Neaj}\ \emph {et~al.}(2020)\citenamefont {El-Neaj}
  \emph {et~al.}}]{Bertoldi:2019tck}%
  \BibitemOpen
  \bibfield  {author} {\bibinfo {author} {\bibfnamefont {Y.~A.}\ \bibnamefont
  {El-Neaj}} \emph {et~al.} (\bibinfo {collaboration} {AEDGE}),\ }\href
  {\doibase 10.1140/epjqt/s40507-020-0080-0} {\bibfield  {journal} {\bibinfo
  {journal} {EPJ Quant. Technol.}\ }\textbf {\bibinfo {volume} {7}},\ \bibinfo
  {pages} {6} (\bibinfo {year} {2020})},\ \Eprint
  {http://arxiv.org/abs/1908.00802} {arXiv:1908.00802 [gr-qc]} \BibitemShut
  {NoStop}%
\bibitem [{\citenamefont {Yagi}\ and\ \citenamefont
  {Seto}(2011)}]{Yagi:2011wg}%
  \BibitemOpen
  \bibfield  {author} {\bibinfo {author} {\bibfnamefont {K.}~\bibnamefont
  {Yagi}}\ and\ \bibinfo {author} {\bibfnamefont {N.}~\bibnamefont {Seto}},\
  }\href {\doibase 10.1103/PhysRevD.83.044011} {\bibfield  {journal} {\bibinfo
  {journal} {Phys. Rev. D}\ }\textbf {\bibinfo {volume} {83}},\ \bibinfo
  {pages} {044011} (\bibinfo {year} {2011})},\ \bibinfo {note} {[Erratum:
  Phys.Rev.D 95, 109901 (2017)]},\ \Eprint {http://arxiv.org/abs/1101.3940}
  {arXiv:1101.3940 [astro-ph.CO]} \BibitemShut {NoStop}%
\bibitem [{\citenamefont {Sato}\ \emph {et~al.}(2017)\citenamefont {Sato} \emph
  {et~al.}}]{Sato_2017}%
  \BibitemOpen
  \bibfield  {author} {\bibinfo {author} {\bibfnamefont {S.}~\bibnamefont
  {Sato}} \emph {et~al.},\ }\href {\doibase 10.1088/1742-6596/840/1/012010}
  {\bibfield  {journal} {\bibinfo  {journal} {Journal of Physics: Conference
  Series}\ }\textbf {\bibinfo {volume} {840}},\ \bibinfo {pages} {012010}
  (\bibinfo {year} {2017})}\BibitemShut {NoStop}%
\bibitem [{\citenamefont {Janssen}\ \emph {et~al.}(2015)\citenamefont {Janssen}
  \emph {et~al.}}]{Janssen:2014dka}%
  \BibitemOpen
  \bibfield  {author} {\bibinfo {author} {\bibfnamefont {G.}~\bibnamefont
  {Janssen}} \emph {et~al.},\ }\href {\doibase 10.22323/1.215.0037} {\bibfield
  {journal} {\bibinfo  {journal} {PoS}\ }\textbf {\bibinfo {volume}
  {AASKA14}},\ \bibinfo {pages} {037} (\bibinfo {year} {2015})},\ \Eprint
  {http://arxiv.org/abs/1501.00127} {arXiv:1501.00127 [astro-ph.IM]}
  \BibitemShut {NoStop}%
\bibitem [{\citenamefont {{Kuns}}\ \emph {et~al.}(2019)\citenamefont {{Kuns}},
  \citenamefont {{Yu}}, \citenamefont {{Chen}},\ and\ \citenamefont
  {{Adhikari}}}]{TianGO}%
  \BibitemOpen
  \bibfield  {author} {\bibinfo {author} {\bibfnamefont {K.~A.}\ \bibnamefont
  {{Kuns}}}, \bibinfo {author} {\bibfnamefont {H.}~\bibnamefont {{Yu}}},
  \bibinfo {author} {\bibfnamefont {Y.}~\bibnamefont {{Chen}}}, \ and\ \bibinfo
  {author} {\bibfnamefont {R.~X.}\ \bibnamefont {{Adhikari}}},\ }\href@noop {}
  {\bibfield  {journal} {\bibinfo  {journal} {arXiv e-prints}\ } (\bibinfo
  {year} {2019})},\ \Eprint {http://arxiv.org/abs/1908.06004} {arXiv:1908.06004
  [gr-qc]} \BibitemShut {NoStop}%
\bibitem [{\citenamefont {Blas}\ and\ \citenamefont
  {Jenkins}(2021)}]{Blas:2021mqw}%
  \BibitemOpen
  \bibfield  {author} {\bibinfo {author} {\bibfnamefont {D.}~\bibnamefont
  {Blas}}\ and\ \bibinfo {author} {\bibfnamefont {A.~C.}\ \bibnamefont
  {Jenkins}},\ }\href@noop {} {\  (\bibinfo {year} {2021})},\ \Eprint
  {http://arxiv.org/abs/2107.04601} {arXiv:2107.04601 [astro-ph.CO]}
  \BibitemShut {NoStop}%
\bibitem [{\citenamefont {Sesana}\ \emph {et~al.}(2021)\citenamefont {Sesana}
  \emph {et~al.}}]{Sesana:2019vho}%
  \BibitemOpen
  \bibfield  {author} {\bibinfo {author} {\bibfnamefont {A.}~\bibnamefont
  {Sesana}} \emph {et~al.},\ }\href {\doibase 10.1007/s10686-021-09709-9}
  {\bibfield  {journal} {\bibinfo  {journal} {Exper. Astron.}\ }\textbf
  {\bibinfo {volume} {51}},\ \bibinfo {pages} {1333} (\bibinfo {year}
  {2021})},\ \Eprint {http://arxiv.org/abs/1908.11391} {arXiv:1908.11391
  [astro-ph.IM]} \BibitemShut {NoStop}%
\bibitem [{\citenamefont {Linde}(1994)}]{Linde:1993cn}%
  \BibitemOpen
  \bibfield  {author} {\bibinfo {author} {\bibfnamefont {A.~D.}\ \bibnamefont
  {Linde}},\ }\href {\doibase 10.1103/PhysRevD.49.748} {\bibfield  {journal}
  {\bibinfo  {journal} {Phys. Rev. D}\ }\textbf {\bibinfo {volume} {49}},\
  \bibinfo {pages} {748} (\bibinfo {year} {1994})},\ \Eprint
  {http://arxiv.org/abs/astro-ph/9307002} {arXiv:astro-ph/9307002} \BibitemShut
  {NoStop}%
\bibitem [{\citenamefont {Linde}(1991)}]{Linde:1991km}%
  \BibitemOpen
  \bibfield  {author} {\bibinfo {author} {\bibfnamefont {A.~D.}\ \bibnamefont
  {Linde}},\ }\href {\doibase 10.1016/0370-2693(91)90130-I} {\bibfield
  {journal} {\bibinfo  {journal} {Phys. Lett. B}\ }\textbf {\bibinfo {volume}
  {259}},\ \bibinfo {pages} {38} (\bibinfo {year} {1991})}\BibitemShut
  {NoStop}%
\bibitem [{\citenamefont {Garcia-Bellido}\ \emph {et~al.}(1996)\citenamefont
  {Garcia-Bellido}, \citenamefont {Linde},\ and\ \citenamefont
  {Wands}}]{Garcia-Bellido:1996mdl}%
  \BibitemOpen
  \bibfield  {author} {\bibinfo {author} {\bibfnamefont {J.}~\bibnamefont
  {Garcia-Bellido}}, \bibinfo {author} {\bibfnamefont {A.~D.}\ \bibnamefont
  {Linde}}, \ and\ \bibinfo {author} {\bibfnamefont {D.}~\bibnamefont
  {Wands}},\ }\href {\doibase 10.1103/PhysRevD.54.6040} {\bibfield  {journal}
  {\bibinfo  {journal} {Phys. Rev. D}\ }\textbf {\bibinfo {volume} {54}},\
  \bibinfo {pages} {6040} (\bibinfo {year} {1996})},\ \Eprint
  {http://arxiv.org/abs/astro-ph/9605094} {arXiv:astro-ph/9605094} \BibitemShut
  {NoStop}%
\bibitem [{\citenamefont {Linde}\ and\ \citenamefont
  {Riotto}(1997)}]{Linde:1997sj}%
  \BibitemOpen
  \bibfield  {author} {\bibinfo {author} {\bibfnamefont {A.~D.}\ \bibnamefont
  {Linde}}\ and\ \bibinfo {author} {\bibfnamefont {A.}~\bibnamefont {Riotto}},\
  }\href {\doibase 10.1103/PhysRevD.56.R1841} {\bibfield  {journal} {\bibinfo
  {journal} {Phys. Rev. D}\ }\textbf {\bibinfo {volume} {56}},\ \bibinfo
  {pages} {R1841} (\bibinfo {year} {1997})},\ \Eprint
  {http://arxiv.org/abs/hep-ph/9703209} {arXiv:hep-ph/9703209} \BibitemShut
  {NoStop}%
\bibitem [{\citenamefont {Randall}\ \emph {et~al.}(1996)\citenamefont
  {Randall}, \citenamefont {Soljacic},\ and\ \citenamefont
  {Guth}}]{Randall:1995dj}%
  \BibitemOpen
  \bibfield  {author} {\bibinfo {author} {\bibfnamefont {L.}~\bibnamefont
  {Randall}}, \bibinfo {author} {\bibfnamefont {M.}~\bibnamefont {Soljacic}}, \
  and\ \bibinfo {author} {\bibfnamefont {A.~H.}\ \bibnamefont {Guth}},\ }\href
  {\doibase 10.1016/0550-3213(96)00174-5} {\bibfield  {journal} {\bibinfo
  {journal} {Nucl. Phys. B}\ }\textbf {\bibinfo {volume} {472}},\ \bibinfo
  {pages} {377} (\bibinfo {year} {1996})},\ \Eprint
  {http://arxiv.org/abs/hep-ph/9512439} {arXiv:hep-ph/9512439} \BibitemShut
  {NoStop}%
\bibitem [{\citenamefont {Halpern}\ \emph {et~al.}(2015)\citenamefont
  {Halpern}, \citenamefont {Hertzberg}, \citenamefont {Joss},\ and\
  \citenamefont {Sfakianakis}}]{Halpern:2014mca}%
  \BibitemOpen
  \bibfield  {author} {\bibinfo {author} {\bibfnamefont {I.~F.}\ \bibnamefont
  {Halpern}}, \bibinfo {author} {\bibfnamefont {M.~P.}\ \bibnamefont
  {Hertzberg}}, \bibinfo {author} {\bibfnamefont {M.~A.}\ \bibnamefont {Joss}},
  \ and\ \bibinfo {author} {\bibfnamefont {E.~I.}\ \bibnamefont
  {Sfakianakis}},\ }\href {\doibase 10.1016/j.physletb.2015.06.076} {\bibfield
  {journal} {\bibinfo  {journal} {Phys. Lett. B}\ }\textbf {\bibinfo {volume}
  {748}},\ \bibinfo {pages} {132} (\bibinfo {year} {2015})},\ \Eprint
  {http://arxiv.org/abs/1410.1878} {arXiv:1410.1878 [astro-ph.CO]} \BibitemShut
  {NoStop}%
\bibitem [{\citenamefont {Clesse}\ and\ \citenamefont
  {Garc\'\i{}a-Bellido}(2015)}]{Clesse:2015wea}%
  \BibitemOpen
  \bibfield  {author} {\bibinfo {author} {\bibfnamefont {S.}~\bibnamefont
  {Clesse}}\ and\ \bibinfo {author} {\bibfnamefont {J.}~\bibnamefont
  {Garc\'\i{}a-Bellido}},\ }\href {\doibase 10.1103/PhysRevD.92.023524}
  {\bibfield  {journal} {\bibinfo  {journal} {Phys. Rev. D}\ }\textbf {\bibinfo
  {volume} {92}},\ \bibinfo {pages} {023524} (\bibinfo {year} {2015})},\
  \Eprint {http://arxiv.org/abs/1501.07565} {arXiv:1501.07565 [astro-ph.CO]}
  \BibitemShut {NoStop}%
\bibitem [{\citenamefont {Guth}\ and\ \citenamefont
  {Sfakianakis}(2012)}]{Guth:2012we}%
  \BibitemOpen
  \bibfield  {author} {\bibinfo {author} {\bibfnamefont {A.~H.}\ \bibnamefont
  {Guth}}\ and\ \bibinfo {author} {\bibfnamefont {E.~I.}\ \bibnamefont
  {Sfakianakis}},\ }\href@noop {} {\  (\bibinfo {year} {2012})},\ \Eprint
  {http://arxiv.org/abs/1210.8128} {arXiv:1210.8128 [astro-ph.CO]} \BibitemShut
  {NoStop}%
\bibitem [{\citenamefont {Wang}\ and\ \citenamefont
  {Xianyu}(2020)}]{Wang:2020ioa}%
  \BibitemOpen
  \bibfield  {author} {\bibinfo {author} {\bibfnamefont {L.-T.}\ \bibnamefont
  {Wang}}\ and\ \bibinfo {author} {\bibfnamefont {Z.-Z.}\ \bibnamefont
  {Xianyu}},\ }\href {\doibase 10.1007/JHEP11(2020)082} {\bibfield  {journal}
  {\bibinfo  {journal} {JHEP}\ }\textbf {\bibinfo {volume} {11}},\ \bibinfo
  {pages} {082} (\bibinfo {year} {2020})},\ \Eprint
  {http://arxiv.org/abs/2004.02887} {arXiv:2004.02887 [hep-ph]} \BibitemShut
  {NoStop}%
\bibitem [{\citenamefont {Adshead}\ \emph {et~al.}(2015)\citenamefont
  {Adshead}, \citenamefont {Giblin}, \citenamefont {Scully},\ and\
  \citenamefont {Sfakianakis}}]{Adshead:2015pva}%
  \BibitemOpen
  \bibfield  {author} {\bibinfo {author} {\bibfnamefont {P.}~\bibnamefont
  {Adshead}}, \bibinfo {author} {\bibfnamefont {J.~T.}\ \bibnamefont {Giblin}},
  \bibinfo {author} {\bibfnamefont {T.~R.}\ \bibnamefont {Scully}}, \ and\
  \bibinfo {author} {\bibfnamefont {E.~I.}\ \bibnamefont {Sfakianakis}},\
  }\href {\doibase 10.1088/1475-7516/2015/12/034} {\bibfield  {journal}
  {\bibinfo  {journal} {JCAP}\ }\textbf {\bibinfo {volume} {12}},\ \bibinfo
  {pages} {034} (\bibinfo {year} {2015})},\ \Eprint
  {http://arxiv.org/abs/1502.06506} {arXiv:1502.06506 [astro-ph.CO]}
  \BibitemShut {NoStop}%
\bibitem [{\citenamefont {Barnaby}\ \emph {et~al.}(2011)\citenamefont
  {Barnaby}, \citenamefont {Namba},\ and\ \citenamefont
  {Peloso}}]{Barnaby:2011vw}%
  \BibitemOpen
  \bibfield  {author} {\bibinfo {author} {\bibfnamefont {N.}~\bibnamefont
  {Barnaby}}, \bibinfo {author} {\bibfnamefont {R.}~\bibnamefont {Namba}}, \
  and\ \bibinfo {author} {\bibfnamefont {M.}~\bibnamefont {Peloso}},\ }\href
  {\doibase 10.1088/1475-7516/2011/04/009} {\bibfield  {journal} {\bibinfo
  {journal} {JCAP}\ }\textbf {\bibinfo {volume} {04}},\ \bibinfo {pages} {009}
  (\bibinfo {year} {2011})},\ \Eprint {http://arxiv.org/abs/1102.4333}
  {arXiv:1102.4333 [astro-ph.CO]} \BibitemShut {NoStop}%
\bibitem [{\citenamefont {Barnaby}\ \emph {et~al.}(2012)\citenamefont
  {Barnaby}, \citenamefont {Pajer},\ and\ \citenamefont
  {Peloso}}]{Barnaby:2011qe}%
  \BibitemOpen
  \bibfield  {author} {\bibinfo {author} {\bibfnamefont {N.}~\bibnamefont
  {Barnaby}}, \bibinfo {author} {\bibfnamefont {E.}~\bibnamefont {Pajer}}, \
  and\ \bibinfo {author} {\bibfnamefont {M.}~\bibnamefont {Peloso}},\ }\href
  {\doibase 10.1103/PhysRevD.85.023525} {\bibfield  {journal} {\bibinfo
  {journal} {Phys. Rev. D}\ }\textbf {\bibinfo {volume} {85}},\ \bibinfo
  {pages} {023525} (\bibinfo {year} {2012})},\ \Eprint
  {http://arxiv.org/abs/1110.3327} {arXiv:1110.3327 [astro-ph.CO]} \BibitemShut
  {NoStop}%
\bibitem [{\citenamefont {Armendariz-Picon}\ \emph {et~al.}(2008)\citenamefont
  {Armendariz-Picon}, \citenamefont {Trodden},\ and\ \citenamefont
  {West}}]{Armendariz-Picon:2007gbe}%
  \BibitemOpen
  \bibfield  {author} {\bibinfo {author} {\bibfnamefont {C.}~\bibnamefont
  {Armendariz-Picon}}, \bibinfo {author} {\bibfnamefont {M.}~\bibnamefont
  {Trodden}}, \ and\ \bibinfo {author} {\bibfnamefont {E.~J.}\ \bibnamefont
  {West}},\ }\href {\doibase 10.1088/1475-7516/2008/04/036} {\bibfield
  {journal} {\bibinfo  {journal} {JCAP}\ }\textbf {\bibinfo {volume} {04}},\
  \bibinfo {pages} {036} (\bibinfo {year} {2008})},\ \Eprint
  {http://arxiv.org/abs/0707.2177} {arXiv:0707.2177 [hep-ph]} \BibitemShut
  {NoStop}%
\bibitem [{\citenamefont {Dufaux}\ \emph {et~al.}(2006)\citenamefont {Dufaux},
  \citenamefont {Felder}, \citenamefont {Kofman}, \citenamefont {Peloso},\ and\
  \citenamefont {Podolsky}}]{Dufaux:2006ee}%
  \BibitemOpen
  \bibfield  {author} {\bibinfo {author} {\bibfnamefont {J.~F.}\ \bibnamefont
  {Dufaux}}, \bibinfo {author} {\bibfnamefont {G.~N.}\ \bibnamefont {Felder}},
  \bibinfo {author} {\bibfnamefont {L.}~\bibnamefont {Kofman}}, \bibinfo
  {author} {\bibfnamefont {M.}~\bibnamefont {Peloso}}, \ and\ \bibinfo {author}
  {\bibfnamefont {D.}~\bibnamefont {Podolsky}},\ }\href {\doibase
  10.1088/1475-7516/2006/07/006} {\bibfield  {journal} {\bibinfo  {journal}
  {JCAP}\ }\textbf {\bibinfo {volume} {07}},\ \bibinfo {pages} {006} (\bibinfo
  {year} {2006})},\ \Eprint {http://arxiv.org/abs/hep-ph/0602144}
  {arXiv:hep-ph/0602144} \BibitemShut {NoStop}%
\bibitem [{\citenamefont {Garretson}\ \emph {et~al.}(1992)\citenamefont
  {Garretson}, \citenamefont {Field},\ and\ \citenamefont
  {Carroll}}]{Garretson:1992vt}%
  \BibitemOpen
  \bibfield  {author} {\bibinfo {author} {\bibfnamefont {W.~D.}\ \bibnamefont
  {Garretson}}, \bibinfo {author} {\bibfnamefont {G.~B.}\ \bibnamefont
  {Field}}, \ and\ \bibinfo {author} {\bibfnamefont {S.~M.}\ \bibnamefont
  {Carroll}},\ }\href {\doibase 10.1103/PhysRevD.46.5346} {\bibfield  {journal}
  {\bibinfo  {journal} {Phys. Rev. D}\ }\textbf {\bibinfo {volume} {46}},\
  \bibinfo {pages} {5346} (\bibinfo {year} {1992})},\ \Eprint
  {http://arxiv.org/abs/hep-ph/9209238} {arXiv:hep-ph/9209238} \BibitemShut
  {NoStop}%
\bibitem [{\citenamefont {Giblin}\ and\ \citenamefont
  {Thrane}(2014)}]{Giblin:2014gra}%
  \BibitemOpen
  \bibfield  {author} {\bibinfo {author} {\bibfnamefont {J.~T.}\ \bibnamefont
  {Giblin}}\ and\ \bibinfo {author} {\bibfnamefont {E.}~\bibnamefont
  {Thrane}},\ }\href {\doibase 10.1103/PhysRevD.90.107502} {\bibfield
  {journal} {\bibinfo  {journal} {Phys. Rev. D}\ }\textbf {\bibinfo {volume}
  {90}},\ \bibinfo {pages} {107502} (\bibinfo {year} {2014})},\ \Eprint
  {http://arxiv.org/abs/1410.4779} {arXiv:1410.4779 [gr-qc]} \BibitemShut
  {NoStop}%
\bibitem [{\citenamefont {Poulin}\ \emph {et~al.}(2017)\citenamefont {Poulin},
  \citenamefont {Serpico}, \citenamefont {Calore}, \citenamefont {Clesse},\
  and\ \citenamefont {Kohri}}]{Poulin:2017bwe}%
  \BibitemOpen
  \bibfield  {author} {\bibinfo {author} {\bibfnamefont {V.}~\bibnamefont
  {Poulin}}, \bibinfo {author} {\bibfnamefont {P.~D.}\ \bibnamefont {Serpico}},
  \bibinfo {author} {\bibfnamefont {F.}~\bibnamefont {Calore}}, \bibinfo
  {author} {\bibfnamefont {S.}~\bibnamefont {Clesse}}, \ and\ \bibinfo {author}
  {\bibfnamefont {K.}~\bibnamefont {Kohri}},\ }\href {\doibase
  10.1103/PhysRevD.96.083524} {\bibfield  {journal} {\bibinfo  {journal} {Phys.
  Rev. D}\ }\textbf {\bibinfo {volume} {96}},\ \bibinfo {pages} {083524}
  (\bibinfo {year} {2017})},\ \Eprint {http://arxiv.org/abs/1707.04206}
  {arXiv:1707.04206 [astro-ph.CO]} \BibitemShut {NoStop}%
\bibitem [{\citenamefont {Adshead}\ \emph {et~al.}(2016)\citenamefont
  {Adshead}, \citenamefont {Giblin}, \citenamefont {Scully},\ and\
  \citenamefont {Sfakianakis}}]{Adshead:2016iae}%
  \BibitemOpen
  \bibfield  {author} {\bibinfo {author} {\bibfnamefont {P.}~\bibnamefont
  {Adshead}}, \bibinfo {author} {\bibfnamefont {J.~T.}\ \bibnamefont {Giblin}},
  \bibinfo {author} {\bibfnamefont {T.~R.}\ \bibnamefont {Scully}}, \ and\
  \bibinfo {author} {\bibfnamefont {E.~I.}\ \bibnamefont {Sfakianakis}},\
  }\href {\doibase 10.1088/1475-7516/2016/10/039} {\bibfield  {journal}
  {\bibinfo  {journal} {JCAP}\ }\textbf {\bibinfo {volume} {10}},\ \bibinfo
  {pages} {039} (\bibinfo {year} {2016})},\ \Eprint
  {http://arxiv.org/abs/1606.08474} {arXiv:1606.08474 [astro-ph.CO]}
  \BibitemShut {NoStop}%
\bibitem [{\citenamefont {Okano}\ and\ \citenamefont
  {Fujita}(2021)}]{Okano:2020uyr}%
  \BibitemOpen
  \bibfield  {author} {\bibinfo {author} {\bibfnamefont {S.}~\bibnamefont
  {Okano}}\ and\ \bibinfo {author} {\bibfnamefont {T.}~\bibnamefont {Fujita}},\
  }\href {\doibase 10.1088/1475-7516/2021/03/026} {\bibfield  {journal}
  {\bibinfo  {journal} {JCAP}\ }\textbf {\bibinfo {volume} {03}},\ \bibinfo
  {pages} {026} (\bibinfo {year} {2021})},\ \Eprint
  {http://arxiv.org/abs/2005.13833} {arXiv:2005.13833 [astro-ph.CO]}
  \BibitemShut {NoStop}%
\bibitem [{\citenamefont {Katz}\ \emph {et~al.}(2018)\citenamefont {Katz},
  \citenamefont {Kopp}, \citenamefont {Sibiryakov},\ and\ \citenamefont
  {Xue}}]{Katz:2018zrn}%
  \BibitemOpen
  \bibfield  {author} {\bibinfo {author} {\bibfnamefont {A.}~\bibnamefont
  {Katz}}, \bibinfo {author} {\bibfnamefont {J.}~\bibnamefont {Kopp}}, \bibinfo
  {author} {\bibfnamefont {S.}~\bibnamefont {Sibiryakov}}, \ and\ \bibinfo
  {author} {\bibfnamefont {W.}~\bibnamefont {Xue}},\ }\href {\doibase
  10.1088/1475-7516/2018/12/005} {\bibfield  {journal} {\bibinfo  {journal}
  {JCAP}\ }\textbf {\bibinfo {volume} {12}},\ \bibinfo {pages} {005} (\bibinfo
  {year} {2018})},\ \Eprint {http://arxiv.org/abs/1807.11495} {arXiv:1807.11495
  [astro-ph.CO]} \BibitemShut {NoStop}%
\bibitem [{\citenamefont {Smyth}\ \emph {et~al.}(2020)\citenamefont {Smyth},
  \citenamefont {Profumo}, \citenamefont {English}, \citenamefont {Jeltema},
  \citenamefont {McKinnon},\ and\ \citenamefont
  {Guhathakurta}}]{Smyth:2019whb}%
  \BibitemOpen
  \bibfield  {author} {\bibinfo {author} {\bibfnamefont {N.}~\bibnamefont
  {Smyth}}, \bibinfo {author} {\bibfnamefont {S.}~\bibnamefont {Profumo}},
  \bibinfo {author} {\bibfnamefont {S.}~\bibnamefont {English}}, \bibinfo
  {author} {\bibfnamefont {T.}~\bibnamefont {Jeltema}}, \bibinfo {author}
  {\bibfnamefont {K.}~\bibnamefont {McKinnon}}, \ and\ \bibinfo {author}
  {\bibfnamefont {P.}~\bibnamefont {Guhathakurta}},\ }\href {\doibase
  10.1103/PhysRevD.101.063005} {\bibfield  {journal} {\bibinfo  {journal}
  {Phys. Rev. D}\ }\textbf {\bibinfo {volume} {101}},\ \bibinfo {pages}
  {063005} (\bibinfo {year} {2020})},\ \Eprint
  {http://arxiv.org/abs/1910.01285} {arXiv:1910.01285 [astro-ph.CO]}
  \BibitemShut {NoStop}%
\bibitem [{\citenamefont {Takhistov}(2021)}]{Takhistov:2020ssb}%
  \BibitemOpen
  \bibfield  {author} {\bibinfo {author} {\bibfnamefont {V.}~\bibnamefont
  {Takhistov}},\ }\href {\doibase 10.22323/1.390.0610} {\bibfield  {journal}
  {\bibinfo  {journal} {PoS}\ }\textbf {\bibinfo {volume} {ICHEP2020}},\
  \bibinfo {pages} {610} (\bibinfo {year} {2021})}\BibitemShut {NoStop}%
\bibitem [{\citenamefont {Villanueva-Domingo}\ \emph
  {et~al.}(2021)\citenamefont {Villanueva-Domingo}, \citenamefont {Mena},\ and\
  \citenamefont {Palomares-Ruiz}}]{Villanueva-Domingo:2021spv}%
  \BibitemOpen
  \bibfield  {author} {\bibinfo {author} {\bibfnamefont {P.}~\bibnamefont
  {Villanueva-Domingo}}, \bibinfo {author} {\bibfnamefont {O.}~\bibnamefont
  {Mena}}, \ and\ \bibinfo {author} {\bibfnamefont {S.}~\bibnamefont
  {Palomares-Ruiz}},\ }\href {\doibase 10.3389/fspas.2021.681084} {\bibfield
  {journal} {\bibinfo  {journal} {Front. Astron. Space Sci.}\ }\textbf
  {\bibinfo {volume} {8}},\ \bibinfo {pages} {87} (\bibinfo {year} {2021})},\
  \Eprint {http://arxiv.org/abs/2103.12087} {arXiv:2103.12087 [astro-ph.CO]}
  \BibitemShut {NoStop}%
\bibitem [{\citenamefont {Lyth}(2011)}]{Lyth:2010zq}%
  \BibitemOpen
  \bibfield  {author} {\bibinfo {author} {\bibfnamefont {D.~H.}\ \bibnamefont
  {Lyth}},\ }\href {\doibase 10.1088/1475-7516/2011/07/035} {\bibfield
  {journal} {\bibinfo  {journal} {JCAP}\ }\textbf {\bibinfo {volume} {07}},\
  \bibinfo {pages} {035} (\bibinfo {year} {2011})},\ \Eprint
  {http://arxiv.org/abs/1012.4617} {arXiv:1012.4617 [astro-ph.CO]} \BibitemShut
  {NoStop}%
\bibitem [{\citenamefont {Lyth}(2012)}]{Lyth:2012yp}%
  \BibitemOpen
  \bibfield  {author} {\bibinfo {author} {\bibfnamefont {D.~H.}\ \bibnamefont
  {Lyth}},\ }\href {\doibase 10.1088/1475-7516/2012/05/022} {\bibfield
  {journal} {\bibinfo  {journal} {JCAP}\ }\textbf {\bibinfo {volume} {05}},\
  \bibinfo {pages} {022} (\bibinfo {year} {2012})},\ \Eprint
  {http://arxiv.org/abs/1201.4312} {arXiv:1201.4312 [astro-ph.CO]} \BibitemShut
  {NoStop}%
\bibitem [{\citenamefont {Musco}\ \emph {et~al.}(2005)\citenamefont {Musco},
  \citenamefont {Miller},\ and\ \citenamefont {Rezzolla}}]{Musco:2004ak}%
  \BibitemOpen
  \bibfield  {author} {\bibinfo {author} {\bibfnamefont {I.}~\bibnamefont
  {Musco}}, \bibinfo {author} {\bibfnamefont {J.~C.}\ \bibnamefont {Miller}}, \
  and\ \bibinfo {author} {\bibfnamefont {L.}~\bibnamefont {Rezzolla}},\ }\href
  {\doibase 10.1088/0264-9381/22/7/013} {\bibfield  {journal} {\bibinfo
  {journal} {Class. Quant. Grav.}\ }\textbf {\bibinfo {volume} {22}},\ \bibinfo
  {pages} {1405} (\bibinfo {year} {2005})},\ \Eprint
  {http://arxiv.org/abs/gr-qc/0412063} {arXiv:gr-qc/0412063} \BibitemShut
  {NoStop}%
\bibitem [{\citenamefont {Niemeyer}\ and\ \citenamefont
  {Jedamzik}(1999)}]{Niemeyer:1999ak}%
  \BibitemOpen
  \bibfield  {author} {\bibinfo {author} {\bibfnamefont {J.~C.}\ \bibnamefont
  {Niemeyer}}\ and\ \bibinfo {author} {\bibfnamefont {K.}~\bibnamefont
  {Jedamzik}},\ }\href {\doibase 10.1103/PhysRevD.59.124013} {\bibfield
  {journal} {\bibinfo  {journal} {Phys. Rev. D}\ }\textbf {\bibinfo {volume}
  {59}},\ \bibinfo {pages} {124013} (\bibinfo {year} {1999})},\ \Eprint
  {http://arxiv.org/abs/astro-ph/9901292} {arXiv:astro-ph/9901292} \BibitemShut
  {NoStop}%
\bibitem [{\citenamefont {Shibata}\ and\ \citenamefont
  {Sasaki}(1999)}]{Shibata:1999zs}%
  \BibitemOpen
  \bibfield  {author} {\bibinfo {author} {\bibfnamefont {M.}~\bibnamefont
  {Shibata}}\ and\ \bibinfo {author} {\bibfnamefont {M.}~\bibnamefont
  {Sasaki}},\ }\href {\doibase 10.1103/PhysRevD.60.084002} {\bibfield
  {journal} {\bibinfo  {journal} {Phys. Rev. D}\ }\textbf {\bibinfo {volume}
  {60}},\ \bibinfo {pages} {084002} (\bibinfo {year} {1999})},\ \Eprint
  {http://arxiv.org/abs/gr-qc/9905064} {arXiv:gr-qc/9905064} \BibitemShut
  {NoStop}%
\bibitem [{\citenamefont {Yoo}\ \emph {et~al.}(2018)\citenamefont {Yoo},
  \citenamefont {Harada}, \citenamefont {Garriga},\ and\ \citenamefont
  {Kohri}}]{Yoo:2018kvb}%
  \BibitemOpen
  \bibfield  {author} {\bibinfo {author} {\bibfnamefont {C.-M.}\ \bibnamefont
  {Yoo}}, \bibinfo {author} {\bibfnamefont {T.}~\bibnamefont {Harada}},
  \bibinfo {author} {\bibfnamefont {J.}~\bibnamefont {Garriga}}, \ and\
  \bibinfo {author} {\bibfnamefont {K.}~\bibnamefont {Kohri}},\ }\href
  {\doibase 10.1093/ptep/pty120} {\bibfield  {journal} {\bibinfo  {journal}
  {PTEP}\ }\textbf {\bibinfo {volume} {2018}},\ \bibinfo {pages} {123E01}
  (\bibinfo {year} {2018})},\ \Eprint {http://arxiv.org/abs/1805.03946}
  {arXiv:1805.03946 [astro-ph.CO]} \BibitemShut {NoStop}%
\bibitem [{\citenamefont {Yoo}\ \emph {et~al.}(2020)\citenamefont {Yoo},
  \citenamefont {Harada},\ and\ \citenamefont {Okawa}}]{Yoo:2020lmg}%
  \BibitemOpen
  \bibfield  {author} {\bibinfo {author} {\bibfnamefont {C.-M.}\ \bibnamefont
  {Yoo}}, \bibinfo {author} {\bibfnamefont {T.}~\bibnamefont {Harada}}, \ and\
  \bibinfo {author} {\bibfnamefont {H.}~\bibnamefont {Okawa}},\ }\href
  {\doibase 10.1103/PhysRevD.102.043526} {\bibfield  {journal} {\bibinfo
  {journal} {Phys. Rev. D}\ }\textbf {\bibinfo {volume} {102}},\ \bibinfo
  {pages} {043526} (\bibinfo {year} {2020})},\ \Eprint
  {http://arxiv.org/abs/2004.01042} {arXiv:2004.01042 [gr-qc]} \BibitemShut
  {NoStop}%
\bibitem [{\citenamefont {Harada}\ \emph {et~al.}(2013)\citenamefont {Harada},
  \citenamefont {Yoo},\ and\ \citenamefont {Kohri}}]{Harada:2013epa}%
  \BibitemOpen
  \bibfield  {author} {\bibinfo {author} {\bibfnamefont {T.}~\bibnamefont
  {Harada}}, \bibinfo {author} {\bibfnamefont {C.-M.}\ \bibnamefont {Yoo}}, \
  and\ \bibinfo {author} {\bibfnamefont {K.}~\bibnamefont {Kohri}},\ }\href
  {\doibase 10.1103/PhysRevD.88.084051} {\bibfield  {journal} {\bibinfo
  {journal} {Phys. Rev. D}\ }\textbf {\bibinfo {volume} {88}},\ \bibinfo
  {pages} {084051} (\bibinfo {year} {2013})},\ \bibinfo {note} {[Erratum:
  Phys.Rev.D 89, 029903 (2014)]},\ \Eprint {http://arxiv.org/abs/1309.4201}
  {arXiv:1309.4201 [astro-ph.CO]} \BibitemShut {NoStop}%
\bibitem [{\citenamefont {Bloomfield}\ \emph {et~al.}(2016)\citenamefont
  {Bloomfield}, \citenamefont {Face}, \citenamefont {Guth}, \citenamefont
  {Kalia}, \citenamefont {Lam},\ and\ \citenamefont
  {Moss}}]{Bloomfield:2016civ}%
  \BibitemOpen
  \bibfield  {author} {\bibinfo {author} {\bibfnamefont {J.~K.}\ \bibnamefont
  {Bloomfield}}, \bibinfo {author} {\bibfnamefont {S.~H.~P.}\ \bibnamefont
  {Face}}, \bibinfo {author} {\bibfnamefont {A.~H.}\ \bibnamefont {Guth}},
  \bibinfo {author} {\bibfnamefont {S.}~\bibnamefont {Kalia}}, \bibinfo
  {author} {\bibfnamefont {C.}~\bibnamefont {Lam}}, \ and\ \bibinfo {author}
  {\bibfnamefont {Z.}~\bibnamefont {Moss}},\ }\href@noop {} {\  (\bibinfo
  {year} {2016})},\ \Eprint {http://arxiv.org/abs/1612.03890} {arXiv:1612.03890
  [math-ph]} \BibitemShut {NoStop}%
\bibitem [{\citenamefont {Bloomfield}\ \emph {et~al.}(2018)\citenamefont
  {Bloomfield}, \citenamefont {Face}, \citenamefont {Guth}, \citenamefont
  {Kalia},\ and\ \citenamefont {Moss}}]{Bloomfield:2018oku}%
  \BibitemOpen
  \bibfield  {author} {\bibinfo {author} {\bibfnamefont {J.~K.}\ \bibnamefont
  {Bloomfield}}, \bibinfo {author} {\bibfnamefont {S.~H.~P.}\ \bibnamefont
  {Face}}, \bibinfo {author} {\bibfnamefont {A.~H.}\ \bibnamefont {Guth}},
  \bibinfo {author} {\bibfnamefont {S.}~\bibnamefont {Kalia}}, \ and\ \bibinfo
  {author} {\bibfnamefont {Z.}~\bibnamefont {Moss}},\ }\href@noop {} {\
  (\bibinfo {year} {2018})},\ \Eprint {http://arxiv.org/abs/1810.02078}
  {arXiv:1810.02078 [math-ph]} \BibitemShut {NoStop}%
\bibitem [{\citenamefont {Domcke}\ \emph {et~al.}(2020)\citenamefont {Domcke},
  \citenamefont {Garcia-Bellido}, \citenamefont {Peloso}, \citenamefont
  {Pieroni}, \citenamefont {Ricciardone}, \citenamefont {Sorbo},\ and\
  \citenamefont {Tasinato}}]{Domcke:2019zls}%
  \BibitemOpen
  \bibfield  {author} {\bibinfo {author} {\bibfnamefont {V.}~\bibnamefont
  {Domcke}}, \bibinfo {author} {\bibfnamefont {J.}~\bibnamefont
  {Garcia-Bellido}}, \bibinfo {author} {\bibfnamefont {M.}~\bibnamefont
  {Peloso}}, \bibinfo {author} {\bibfnamefont {M.}~\bibnamefont {Pieroni}},
  \bibinfo {author} {\bibfnamefont {A.}~\bibnamefont {Ricciardone}}, \bibinfo
  {author} {\bibfnamefont {L.}~\bibnamefont {Sorbo}}, \ and\ \bibinfo {author}
  {\bibfnamefont {G.}~\bibnamefont {Tasinato}},\ }\href {\doibase
  10.1088/1475-7516/2020/05/028} {\bibfield  {journal} {\bibinfo  {journal}
  {JCAP}\ }\textbf {\bibinfo {volume} {05}},\ \bibinfo {pages} {028} (\bibinfo
  {year} {2020})},\ \Eprint {http://arxiv.org/abs/1910.08052} {arXiv:1910.08052
  [astro-ph.CO]} \BibitemShut {NoStop}%
\bibitem [{\citenamefont {Smith}\ and\ \citenamefont
  {Caldwell}(2017)}]{Smith:2016jqs}%
  \BibitemOpen
  \bibfield  {author} {\bibinfo {author} {\bibfnamefont {T.~L.}\ \bibnamefont
  {Smith}}\ and\ \bibinfo {author} {\bibfnamefont {R.}~\bibnamefont
  {Caldwell}},\ }\href {\doibase 10.1103/PhysRevD.95.044036} {\bibfield
  {journal} {\bibinfo  {journal} {Phys. Rev. D}\ }\textbf {\bibinfo {volume}
  {95}},\ \bibinfo {pages} {044036} (\bibinfo {year} {2017})},\ \Eprint
  {http://arxiv.org/abs/1609.05901} {arXiv:1609.05901 [gr-qc]} \BibitemShut
  {NoStop}%
\bibitem [{\citenamefont {Belgacem}\ and\ \citenamefont
  {Kamionkowski}(2020)}]{Belgacem:2020nda}%
  \BibitemOpen
  \bibfield  {author} {\bibinfo {author} {\bibfnamefont {E.}~\bibnamefont
  {Belgacem}}\ and\ \bibinfo {author} {\bibfnamefont {M.}~\bibnamefont
  {Kamionkowski}},\ }\href {\doibase 10.1103/PhysRevD.102.023004} {\bibfield
  {journal} {\bibinfo  {journal} {Phys. Rev. D}\ }\textbf {\bibinfo {volume}
  {102}},\ \bibinfo {pages} {023004} (\bibinfo {year} {2020})},\ \Eprint
  {http://arxiv.org/abs/2004.05480} {arXiv:2004.05480 [astro-ph.CO]}
  \BibitemShut {NoStop}%
\bibitem [{\citenamefont {Grojean}\ and\ \citenamefont
  {Servant}(2007)}]{Grojean:2006bp}%
  \BibitemOpen
  \bibfield  {author} {\bibinfo {author} {\bibfnamefont {C.}~\bibnamefont
  {Grojean}}\ and\ \bibinfo {author} {\bibfnamefont {G.}~\bibnamefont
  {Servant}},\ }\href {\doibase 10.1103/PhysRevD.75.043507} {\bibfield
  {journal} {\bibinfo  {journal} {Phys. Rev. D}\ }\textbf {\bibinfo {volume}
  {75}},\ \bibinfo {pages} {043507} (\bibinfo {year} {2007})},\ \Eprint
  {http://arxiv.org/abs/hep-ph/0607107} {arXiv:hep-ph/0607107} \BibitemShut
  {NoStop}%
\bibitem [{\citenamefont {Caprini}\ \emph {et~al.}(2009)\citenamefont
  {Caprini}, \citenamefont {Durrer},\ and\ \citenamefont
  {Servant}}]{Caprini:2009yp}%
  \BibitemOpen
  \bibfield  {author} {\bibinfo {author} {\bibfnamefont {C.}~\bibnamefont
  {Caprini}}, \bibinfo {author} {\bibfnamefont {R.}~\bibnamefont {Durrer}}, \
  and\ \bibinfo {author} {\bibfnamefont {G.}~\bibnamefont {Servant}},\ }\href
  {\doibase 10.1088/1475-7516/2009/12/024} {\bibfield  {journal} {\bibinfo
  {journal} {JCAP}\ }\textbf {\bibinfo {volume} {12}},\ \bibinfo {pages} {024}
  (\bibinfo {year} {2009})},\ \Eprint {http://arxiv.org/abs/0909.0622}
  {arXiv:0909.0622 [astro-ph.CO]} \BibitemShut {NoStop}%
\bibitem [{\citenamefont {Vachaspati}\ and\ \citenamefont
  {Vilenkin}(1985)}]{Vachaspati:1984gt}%
  \BibitemOpen
  \bibfield  {author} {\bibinfo {author} {\bibfnamefont {T.}~\bibnamefont
  {Vachaspati}}\ and\ \bibinfo {author} {\bibfnamefont {A.}~\bibnamefont
  {Vilenkin}},\ }\href {\doibase 10.1103/PhysRevD.31.3052} {\bibfield
  {journal} {\bibinfo  {journal} {Phys. Rev. D}\ }\textbf {\bibinfo {volume}
  {31}},\ \bibinfo {pages} {3052} (\bibinfo {year} {1985})}\BibitemShut
  {NoStop}%
\bibitem [{\citenamefont {Auclair}\ \emph {et~al.}(2020)\citenamefont {Auclair}
  \emph {et~al.}}]{Auclair:2019wcv}%
  \BibitemOpen
  \bibfield  {author} {\bibinfo {author} {\bibfnamefont {P.}~\bibnamefont
  {Auclair}} \emph {et~al.},\ }\href {\doibase 10.1088/1475-7516/2020/04/034}
  {\bibfield  {journal} {\bibinfo  {journal} {JCAP}\ }\textbf {\bibinfo
  {volume} {04}},\ \bibinfo {pages} {034} (\bibinfo {year} {2020})},\ \Eprint
  {http://arxiv.org/abs/1909.00819} {arXiv:1909.00819 [astro-ph.CO]}
  \BibitemShut {NoStop}%
\bibitem [{\citenamefont {Barish}\ \emph {et~al.}(2021)\citenamefont {Barish},
  \citenamefont {Bird},\ and\ \citenamefont {Cui}}]{Barish:2020vmy}%
  \BibitemOpen
  \bibfield  {author} {\bibinfo {author} {\bibfnamefont {B.~C.}\ \bibnamefont
  {Barish}}, \bibinfo {author} {\bibfnamefont {S.}~\bibnamefont {Bird}}, \ and\
  \bibinfo {author} {\bibfnamefont {Y.}~\bibnamefont {Cui}},\ }\href {\doibase
  10.1103/PhysRevD.103.123541} {\bibfield  {journal} {\bibinfo  {journal}
  {Phys. Rev. D}\ }\textbf {\bibinfo {volume} {103}},\ \bibinfo {pages}
  {123541} (\bibinfo {year} {2021})},\ \Eprint
  {http://arxiv.org/abs/2012.07874} {arXiv:2012.07874 [gr-qc]} \BibitemShut
  {NoStop}%
\bibitem [{\citenamefont {Cui}\ \emph {et~al.}(2018)\citenamefont {Cui},
  \citenamefont {Lewicki}, \citenamefont {Morrissey},\ and\ \citenamefont
  {Wells}}]{Cui:2017ufi}%
  \BibitemOpen
  \bibfield  {author} {\bibinfo {author} {\bibfnamefont {Y.}~\bibnamefont
  {Cui}}, \bibinfo {author} {\bibfnamefont {M.}~\bibnamefont {Lewicki}},
  \bibinfo {author} {\bibfnamefont {D.~E.}\ \bibnamefont {Morrissey}}, \ and\
  \bibinfo {author} {\bibfnamefont {J.~D.}\ \bibnamefont {Wells}},\ }\href
  {\doibase 10.1103/PhysRevD.97.123505} {\bibfield  {journal} {\bibinfo
  {journal} {Phys. Rev. D}\ }\textbf {\bibinfo {volume} {97}},\ \bibinfo
  {pages} {123505} (\bibinfo {year} {2018})},\ \Eprint
  {http://arxiv.org/abs/1711.03104} {arXiv:1711.03104 [hep-ph]} \BibitemShut
  {NoStop}%
\bibitem [{\citenamefont {Cui}\ \emph {et~al.}(2019)\citenamefont {Cui},
  \citenamefont {Lewicki}, \citenamefont {Morrissey},\ and\ \citenamefont
  {Wells}}]{Cui:2018rwi}%
  \BibitemOpen
  \bibfield  {author} {\bibinfo {author} {\bibfnamefont {Y.}~\bibnamefont
  {Cui}}, \bibinfo {author} {\bibfnamefont {M.}~\bibnamefont {Lewicki}},
  \bibinfo {author} {\bibfnamefont {D.~E.}\ \bibnamefont {Morrissey}}, \ and\
  \bibinfo {author} {\bibfnamefont {J.~D.}\ \bibnamefont {Wells}},\ }\href
  {\doibase 10.1007/JHEP01(2019)081} {\bibfield  {journal} {\bibinfo  {journal}
  {JHEP}\ }\textbf {\bibinfo {volume} {01}},\ \bibinfo {pages} {081} (\bibinfo
  {year} {2019})},\ \Eprint {http://arxiv.org/abs/1808.08968} {arXiv:1808.08968
  [hep-ph]} \BibitemShut {NoStop}%
\bibitem [{\citenamefont {Caprini}\ and\ \citenamefont
  {Figueroa}(2018)}]{Caprini:2018mtu}%
  \BibitemOpen
  \bibfield  {author} {\bibinfo {author} {\bibfnamefont {C.}~\bibnamefont
  {Caprini}}\ and\ \bibinfo {author} {\bibfnamefont {D.~G.}\ \bibnamefont
  {Figueroa}},\ }\href {\doibase 10.1088/1361-6382/aac608} {\bibfield
  {journal} {\bibinfo  {journal} {Class. Quant. Grav.}\ }\textbf {\bibinfo
  {volume} {35}},\ \bibinfo {pages} {163001} (\bibinfo {year} {2018})},\
  \Eprint {http://arxiv.org/abs/1801.04268} {arXiv:1801.04268 [astro-ph.CO]}
  \BibitemShut {NoStop}%
\bibitem [{\citenamefont {Caprini}\ \emph {et~al.}(2020)\citenamefont {Caprini}
  \emph {et~al.}}]{Caprini:2019egz}%
  \BibitemOpen
  \bibfield  {author} {\bibinfo {author} {\bibfnamefont {C.}~\bibnamefont
  {Caprini}} \emph {et~al.},\ }\href {\doibase 10.1088/1475-7516/2020/03/024}
  {\bibfield  {journal} {\bibinfo  {journal} {JCAP}\ }\textbf {\bibinfo
  {volume} {03}},\ \bibinfo {pages} {024} (\bibinfo {year} {2020})},\ \Eprint
  {http://arxiv.org/abs/1910.13125} {arXiv:1910.13125 [astro-ph.CO]}
  \BibitemShut {NoStop}%
\bibitem [{\citenamefont {Thorne}\ \emph {et~al.}(2018)\citenamefont {Thorne},
  \citenamefont {Fujita}, \citenamefont {Hazumi}, \citenamefont {Katayama},
  \citenamefont {Komatsu},\ and\ \citenamefont {Shiraishi}}]{Thorne:2017jft}%
  \BibitemOpen
  \bibfield  {author} {\bibinfo {author} {\bibfnamefont {B.}~\bibnamefont
  {Thorne}}, \bibinfo {author} {\bibfnamefont {T.}~\bibnamefont {Fujita}},
  \bibinfo {author} {\bibfnamefont {M.}~\bibnamefont {Hazumi}}, \bibinfo
  {author} {\bibfnamefont {N.}~\bibnamefont {Katayama}}, \bibinfo {author}
  {\bibfnamefont {E.}~\bibnamefont {Komatsu}}, \ and\ \bibinfo {author}
  {\bibfnamefont {M.}~\bibnamefont {Shiraishi}},\ }\href {\doibase
  10.1103/PhysRevD.97.043506} {\bibfield  {journal} {\bibinfo  {journal} {Phys.
  Rev. D}\ }\textbf {\bibinfo {volume} {97}},\ \bibinfo {pages} {043506}
  (\bibinfo {year} {2018})},\ \Eprint {http://arxiv.org/abs/1707.03240}
  {arXiv:1707.03240 [astro-ph.CO]} \BibitemShut {NoStop}%
\bibitem [{\citenamefont {Caldwell}\ and\ \citenamefont
  {Devulder}(2018)}]{Caldwell:2017chz}%
  \BibitemOpen
  \bibfield  {author} {\bibinfo {author} {\bibfnamefont {R.~R.}\ \bibnamefont
  {Caldwell}}\ and\ \bibinfo {author} {\bibfnamefont {C.}~\bibnamefont
  {Devulder}},\ }\href {\doibase 10.1103/PhysRevD.97.023532} {\bibfield
  {journal} {\bibinfo  {journal} {Phys. Rev. D}\ }\textbf {\bibinfo {volume}
  {97}},\ \bibinfo {pages} {023532} (\bibinfo {year} {2018})},\ \Eprint
  {http://arxiv.org/abs/1706.03765} {arXiv:1706.03765 [astro-ph.CO]}
  \BibitemShut {NoStop}%
\bibitem [{\citenamefont {Dimastrogiovanni}\ \emph {et~al.}(2017)\citenamefont
  {Dimastrogiovanni}, \citenamefont {Fasiello},\ and\ \citenamefont
  {Fujita}}]{Dimastrogiovanni:2016fuu}%
  \BibitemOpen
  \bibfield  {author} {\bibinfo {author} {\bibfnamefont {E.}~\bibnamefont
  {Dimastrogiovanni}}, \bibinfo {author} {\bibfnamefont {M.}~\bibnamefont
  {Fasiello}}, \ and\ \bibinfo {author} {\bibfnamefont {T.}~\bibnamefont
  {Fujita}},\ }\href {\doibase 10.1088/1475-7516/2017/01/019} {\bibfield
  {journal} {\bibinfo  {journal} {JCAP}\ }\textbf {\bibinfo {volume} {01}},\
  \bibinfo {pages} {019} (\bibinfo {year} {2017})},\ \Eprint
  {http://arxiv.org/abs/1608.04216} {arXiv:1608.04216 [astro-ph.CO]}
  \BibitemShut {NoStop}%
\bibitem [{\citenamefont {Fujita}\ \emph {et~al.}(2019)\citenamefont {Fujita},
  \citenamefont {Sfakianakis},\ and\ \citenamefont
  {Shiraishi}}]{Fujita:2018ndp}%
  \BibitemOpen
  \bibfield  {author} {\bibinfo {author} {\bibfnamefont {T.}~\bibnamefont
  {Fujita}}, \bibinfo {author} {\bibfnamefont {E.~I.}\ \bibnamefont
  {Sfakianakis}}, \ and\ \bibinfo {author} {\bibfnamefont {M.}~\bibnamefont
  {Shiraishi}},\ }\href {\doibase 10.1088/1475-7516/2019/05/057} {\bibfield
  {journal} {\bibinfo  {journal} {JCAP}\ }\textbf {\bibinfo {volume} {05}},\
  \bibinfo {pages} {057} (\bibinfo {year} {2019})},\ \Eprint
  {http://arxiv.org/abs/1812.03667} {arXiv:1812.03667 [astro-ph.CO]}
  \BibitemShut {NoStop}%
\bibitem [{\citenamefont {D'Amico}\ and\ \citenamefont
  {Kaloper}(2021)}]{DAmico:2020euu}%
  \BibitemOpen
  \bibfield  {author} {\bibinfo {author} {\bibfnamefont {G.}~\bibnamefont
  {D'Amico}}\ and\ \bibinfo {author} {\bibfnamefont {N.}~\bibnamefont
  {Kaloper}},\ }\href {\doibase 10.1088/1475-7516/2021/08/058} {\bibfield
  {journal} {\bibinfo  {journal} {JCAP}\ }\textbf {\bibinfo {volume} {08}},\
  \bibinfo {pages} {058} (\bibinfo {year} {2021})},\ \Eprint
  {http://arxiv.org/abs/2011.09489} {arXiv:2011.09489 [hep-th]} \BibitemShut
  {NoStop}%
\bibitem [{\citenamefont {D'Amico}\ \emph {et~al.}(2021)\citenamefont
  {D'Amico}, \citenamefont {Kaloper},\ and\ \citenamefont
  {Westphal}}]{DAmico:2021vka}%
  \BibitemOpen
  \bibfield  {author} {\bibinfo {author} {\bibfnamefont {G.}~\bibnamefont
  {D'Amico}}, \bibinfo {author} {\bibfnamefont {N.}~\bibnamefont {Kaloper}}, \
  and\ \bibinfo {author} {\bibfnamefont {A.}~\bibnamefont {Westphal}},\ }\href
  {\doibase 10.1103/PhysRevD.104.L081302} {\  (\bibinfo {year} {2021}),\
  10.1103/PhysRevD.104.L081302},\ \Eprint {http://arxiv.org/abs/2101.05861}
  {arXiv:2101.05861 [hep-th]} \BibitemShut {NoStop}%
\bibitem [{\citenamefont {Zyla}\ \emph {et~al.}(2020)\citenamefont {Zyla} \emph
  {et~al.}}]{ParticleDataGroup:2020ssz}%
  \BibitemOpen
  \bibfield  {author} {\bibinfo {author} {\bibfnamefont {P.~A.}\ \bibnamefont
  {Zyla}} \emph {et~al.} (\bibinfo {collaboration} {Particle Data Group}),\
  }\href {\doibase 10.1093/ptep/ptaa104} {\bibfield  {journal} {\bibinfo
  {journal} {PTEP}\ }\textbf {\bibinfo {volume} {2020}},\ \bibinfo {pages}
  {083C01} (\bibinfo {year} {2020})}\BibitemShut {NoStop}%
\bibitem [{\citenamefont {Cadamuro}\ \emph {et~al.}(2011)\citenamefont
  {Cadamuro}, \citenamefont {Hannestad}, \citenamefont {Raffelt},\ and\
  \citenamefont {Redondo}}]{Cadamuro:2010cz}%
  \BibitemOpen
  \bibfield  {author} {\bibinfo {author} {\bibfnamefont {D.}~\bibnamefont
  {Cadamuro}}, \bibinfo {author} {\bibfnamefont {S.}~\bibnamefont {Hannestad}},
  \bibinfo {author} {\bibfnamefont {G.}~\bibnamefont {Raffelt}}, \ and\
  \bibinfo {author} {\bibfnamefont {J.}~\bibnamefont {Redondo}},\ }\href
  {\doibase 10.1088/1475-7516/2011/02/003} {\bibfield  {journal} {\bibinfo
  {journal} {JCAP}\ }\textbf {\bibinfo {volume} {02}},\ \bibinfo {pages} {003}
  (\bibinfo {year} {2011})},\ \Eprint {http://arxiv.org/abs/1011.3694}
  {arXiv:1011.3694 [hep-ph]} \BibitemShut {NoStop}%
\bibitem [{\citenamefont {Cadamuro}\ and\ \citenamefont
  {Redondo}(2012)}]{Cadamuro:2011fd}%
  \BibitemOpen
  \bibfield  {author} {\bibinfo {author} {\bibfnamefont {D.}~\bibnamefont
  {Cadamuro}}\ and\ \bibinfo {author} {\bibfnamefont {J.}~\bibnamefont
  {Redondo}},\ }\href {\doibase 10.1088/1475-7516/2012/02/032} {\bibfield
  {journal} {\bibinfo  {journal} {JCAP}\ }\textbf {\bibinfo {volume} {02}},\
  \bibinfo {pages} {032} (\bibinfo {year} {2012})},\ \Eprint
  {http://arxiv.org/abs/1110.2895} {arXiv:1110.2895 [hep-ph]} \BibitemShut
  {NoStop}%
\bibitem [{\citenamefont {Depta}\ \emph {et~al.}(2020)\citenamefont {Depta},
  \citenamefont {Hufnagel},\ and\ \citenamefont
  {Schmidt-Hoberg}}]{Depta:2020wmr}%
  \BibitemOpen
  \bibfield  {author} {\bibinfo {author} {\bibfnamefont {P.~F.}\ \bibnamefont
  {Depta}}, \bibinfo {author} {\bibfnamefont {M.}~\bibnamefont {Hufnagel}}, \
  and\ \bibinfo {author} {\bibfnamefont {K.}~\bibnamefont {Schmidt-Hoberg}},\
  }\href {\doibase 10.1088/1475-7516/2020/05/009} {\bibfield  {journal}
  {\bibinfo  {journal} {JCAP}\ }\textbf {\bibinfo {volume} {05}},\ \bibinfo
  {pages} {009} (\bibinfo {year} {2020})},\ \Eprint
  {http://arxiv.org/abs/2002.08370} {arXiv:2002.08370 [hep-ph]} \BibitemShut
  {NoStop}%
\bibitem [{\citenamefont {Carrasco}\ \emph {et~al.}(2015)\citenamefont
  {Carrasco}, \citenamefont {Kallosh},\ and\ \citenamefont
  {Linde}}]{Carrasco:2015rva}%
  \BibitemOpen
  \bibfield  {author} {\bibinfo {author} {\bibfnamefont {J.~J.~M.}\
  \bibnamefont {Carrasco}}, \bibinfo {author} {\bibfnamefont {R.}~\bibnamefont
  {Kallosh}}, \ and\ \bibinfo {author} {\bibfnamefont {A.}~\bibnamefont
  {Linde}},\ }\href {\doibase 10.1103/PhysRevD.92.063519} {\bibfield  {journal}
  {\bibinfo  {journal} {Phys. Rev. D}\ }\textbf {\bibinfo {volume} {92}},\
  \bibinfo {pages} {063519} (\bibinfo {year} {2015})},\ \Eprint
  {http://arxiv.org/abs/1506.00936} {arXiv:1506.00936 [hep-th]} \BibitemShut
  {NoStop}%
\bibitem [{\citenamefont {Lozanov}\ and\ \citenamefont
  {Amin}(2018)}]{Lozanov:2017hjm}%
  \BibitemOpen
  \bibfield  {author} {\bibinfo {author} {\bibfnamefont {K.~D.}\ \bibnamefont
  {Lozanov}}\ and\ \bibinfo {author} {\bibfnamefont {M.~A.}\ \bibnamefont
  {Amin}},\ }\href {\doibase 10.1103/PhysRevD.97.023533} {\bibfield  {journal}
  {\bibinfo  {journal} {Phys. Rev. D}\ }\textbf {\bibinfo {volume} {97}},\
  \bibinfo {pages} {023533} (\bibinfo {year} {2018})},\ \Eprint
  {http://arxiv.org/abs/1710.06851} {arXiv:1710.06851 [astro-ph.CO]}
  \BibitemShut {NoStop}%
\bibitem [{\citenamefont {Krajewski}\ \emph {et~al.}(2019)\citenamefont
  {Krajewski}, \citenamefont {Turzy\'nski},\ and\ \citenamefont
  {Wieczorek}}]{Krajewski:2018moi}%
  \BibitemOpen
  \bibfield  {author} {\bibinfo {author} {\bibfnamefont {T.}~\bibnamefont
  {Krajewski}}, \bibinfo {author} {\bibfnamefont {K.}~\bibnamefont
  {Turzy\'nski}}, \ and\ \bibinfo {author} {\bibfnamefont {M.}~\bibnamefont
  {Wieczorek}},\ }\href {\doibase 10.1140/epjc/s10052-019-7155-z} {\bibfield
  {journal} {\bibinfo  {journal} {Eur. Phys. J. C}\ }\textbf {\bibinfo {volume}
  {79}},\ \bibinfo {pages} {654} (\bibinfo {year} {2019})},\ \Eprint
  {http://arxiv.org/abs/1801.01786} {arXiv:1801.01786 [astro-ph.CO]}
  \BibitemShut {NoStop}%
\bibitem [{\citenamefont {Iarygina}\ \emph {et~al.}(2020)\citenamefont
  {Iarygina}, \citenamefont {Sfakianakis}, \citenamefont {Wang},\ and\
  \citenamefont {Ach\'ucarro}}]{Iarygina:2020dwe}%
  \BibitemOpen
  \bibfield  {author} {\bibinfo {author} {\bibfnamefont {O.}~\bibnamefont
  {Iarygina}}, \bibinfo {author} {\bibfnamefont {E.~I.}\ \bibnamefont
  {Sfakianakis}}, \bibinfo {author} {\bibfnamefont {D.-G.}\ \bibnamefont
  {Wang}}, \ and\ \bibinfo {author} {\bibfnamefont {A.}~\bibnamefont
  {Ach\'ucarro}},\ }\href@noop {} {\  (\bibinfo {year} {2020})},\ \Eprint
  {http://arxiv.org/abs/2005.00528} {arXiv:2005.00528 [astro-ph.CO]}
  \BibitemShut {NoStop}%
\bibitem [{\citenamefont {Adshead}\ \emph
  {et~al.}(2020{\natexlab{b}})\citenamefont {Adshead}, \citenamefont {Giblin},
  \citenamefont {Pieroni},\ and\ \citenamefont {Weiner}}]{Adshead:2019igv}%
  \BibitemOpen
  \bibfield  {author} {\bibinfo {author} {\bibfnamefont {P.}~\bibnamefont
  {Adshead}}, \bibinfo {author} {\bibfnamefont {J.~T.}\ \bibnamefont {Giblin}},
  \bibinfo {author} {\bibfnamefont {M.}~\bibnamefont {Pieroni}}, \ and\
  \bibinfo {author} {\bibfnamefont {Z.~J.}\ \bibnamefont {Weiner}},\ }\href
  {\doibase 10.1103/PhysRevLett.124.171301} {\bibfield  {journal} {\bibinfo
  {journal} {Phys. Rev. Lett.}\ }\textbf {\bibinfo {volume} {124}},\ \bibinfo
  {pages} {171301} (\bibinfo {year} {2020}{\natexlab{b}})},\ \Eprint
  {http://arxiv.org/abs/1909.12843} {arXiv:1909.12843 [astro-ph.CO]}
  \BibitemShut {NoStop}%
\end{thebibliography}%

\clearpage
\newpage
\onecolumngrid
\setcounter{secnumdepth}{3}
\setcounter{equation}{0}
\setcounter{figure}{0}
\setcounter{table}{0}
\setcounter{page}{1}
\makeatletter
\renewcommand{\theequation}{S\arabic{equation}}
\renewcommand{\thefigure}{S\arabic{figure}}

\begin{center}
\Large{\textbf{Detectable Gravitational Wave Signals from Inflationary Preheating}}\\
\medskip
\textit{Supplemental Material}\\
\medskip
{Yanou Cui and Evangelos I. Sfakianakis}
\end{center}

\vspace{0.5cm}

\section*{why hybrid inflation}
One may question the necessity of a hybrid inflation potential for this mechanism to work and produce observable GW's. A simple alternative for lowering the scale of inflation would be to consider a single-field inflationary sector coupled to a $U(1)$ field. 
The model that we use is
\begin{equation}
S = \int d^4 x \sqrt{-g} \left [
{M_{\rm Pl}^2 \over 2}R -{1\over 2} \partial_\mu \phi \partial^\mu \phi - V(\phi)   -{1\over 4}F_{\mu\nu}F^{\mu\nu}
-{1\over \Lambda}  \phi F_{\mu\nu}\tilde F^{\mu\nu}
\right ]
\end{equation}
where the potential chosen for the inflaton $\phi$ corresponds to the  well-known T-model of $\alpha$-attractors \cite{Carrasco:2015rva}
\beq
V(\phi) = \mu^2 \alpha \tanh \left (
{\phi\over \sqrt{6\alpha}}
\right )^2
\label{eq:Tmodel}
\eeq
While $\alpha$-attractors are by construction two-field models, we neglect the second field and only use the single-field potential of Eq.~\eqref{eq:Tmodel}. The important characteristic of this potential is that the parameter $\alpha$ controls the scale of inflation (and the tensor-to-scalar ratio), while the amplitude of the scalar fluctuations and the spectral index $n_s$ are independent of $\alpha$. In particular $n_s=1-2/N_*$, in agreement  with CMB measurements for proper values of $N_*$. 
The amplitude of scalar fluctuations fixes the mass-scale $\mu \sim 10^{-6}\, M_{\rm Pl}$, regardless of $\alpha$.

The single-field $\alpha$-attractor potential of Eq.~\eqref{eq:Tmodel} exhibits a strong separation of scales for $\alpha\ll1$. The  Hubble scale scales as $H^2 \sim \alpha \mu^2/M_{\rm Pl}^2 $, while the relevant mass-scale during preheating is $\mu$.  The two are only related for $\alpha\sim M_{\rm Pl}$, which is the choice leading to $r={\cal O}(10^{-3})$. Hence, by reducing the scale of inflation, by reducing $\alpha$, the relevant wavenumbers during preheating do not scale as $k\sim H$, but rather as $k\sim \mu\gg H$.

The post-inflationary evolution for the T-model can deviate significantly from the picture with coherently oscillating inflaton. Eq.~\eqref{eq:Tmodel} exhibits strong self-resonance for $\alpha\ll1$ (equivalently $r\ll1$), leading to a quick breakdown of the inflaton condensate  and the emergence of overdensities and oscillons~\cite{Lozanov:2017hjm}.
The full T-model, as derived from supergravity,  consists of an inflaton field with the potential of Eq.~\eqref{eq:Tmodel}, along with a spectator field, which is stabilized during inflation. For $r\ll1$, the spectator field exhibits very efficient preheating, leading to a fast transfer of energy from the inflaton condensate to spectator field fluctuations \cite{Krajewski:2018moi, Iarygina:2020dwe}.

Neglecting the self-resonance of the inflaton and the tachyonic amplification of the spectator field described above, we can compute the amplification of the gauge field after the end of $\alpha$-attractor inflation. 
For $\alpha={\cal O}(1)$, we find that (neglecting non-linearities and back-reaction), the dominant gauge helicity is amplified significantly more and the peak value occurs at $k\simeq \mu$, leading to a very high frequency of GW's, $\nu ={\cal O}(10^7) \, {\rm Hz}$. 
By decreasing the Hubble scale through decreasing the value of $\alpha$, the period of background oscillations remains almost unchanged, as it is controlled by $\mu ={\cal O}(10^{-5}) \,M_{\rm Pl}$. The background field velocity is reduced and in order for parametric resonance to be efficient, we need to increase the Chern-Simmons coupling strength $1/\Lambda$.  The peak of the gauge field power spectrum remains at $k ={\cal O}(\mu)$. Since the frequency of GW's relates to $H$ as $\nu \propto k/\sqrt{H}$, lowering the energy scale of inflation increases the frequency of GW's instead of decreasing it (see also Ref.~\cite{Iarygina:2020dwe}).

\section*{Waterfall and gauge field dynamics: An example in detail}

In order to elucidate our analysis, we explain one of the benchmark examples in detail. We consider the Hubble scale during inflation to be $H=10^{-20}\, {\rm Hz}$, while the waterfall and timer field masses are $m_0=6H$ and $m_\psi=0.5H$ respectively.

\subsection*{Waterfall Dynamics }

\begin{figure}[h!]
\centering
\includegraphics[width=.45\textwidth]{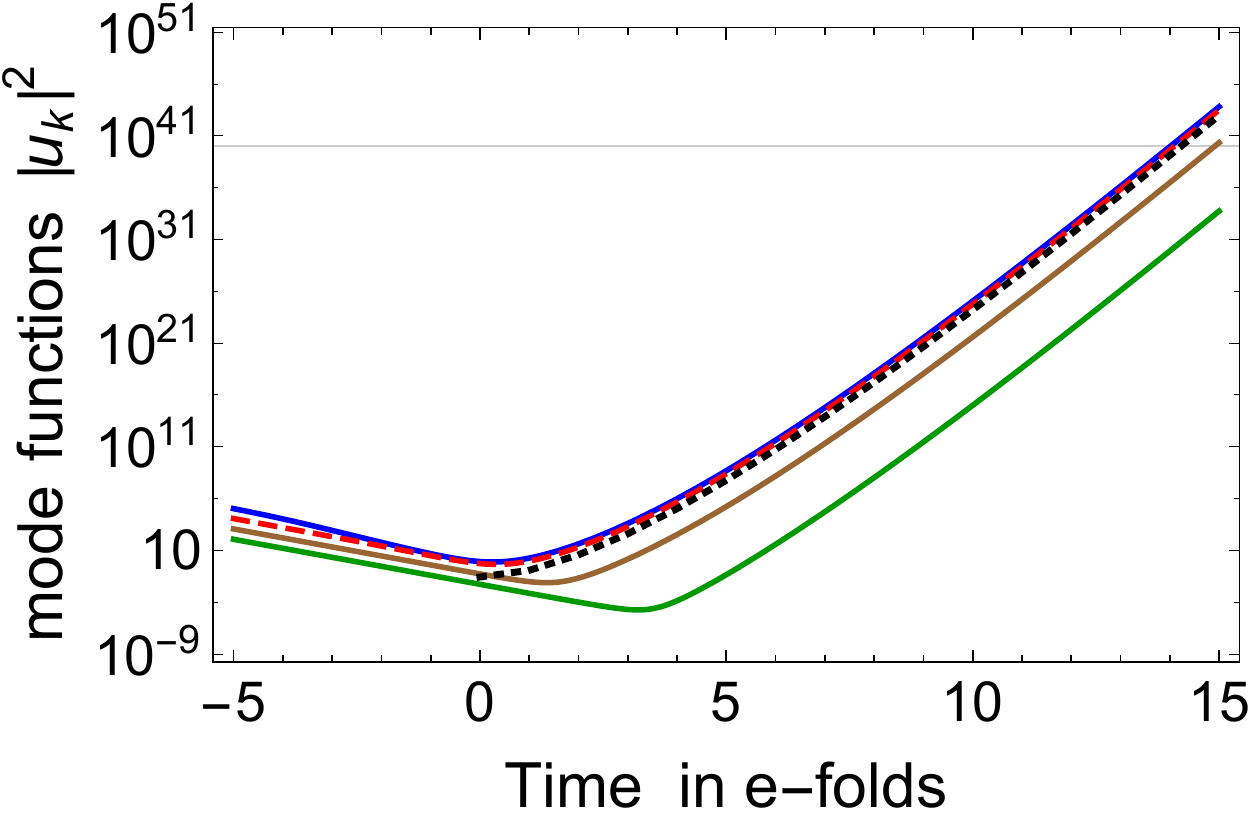}
\includegraphics[width=.45\textwidth]{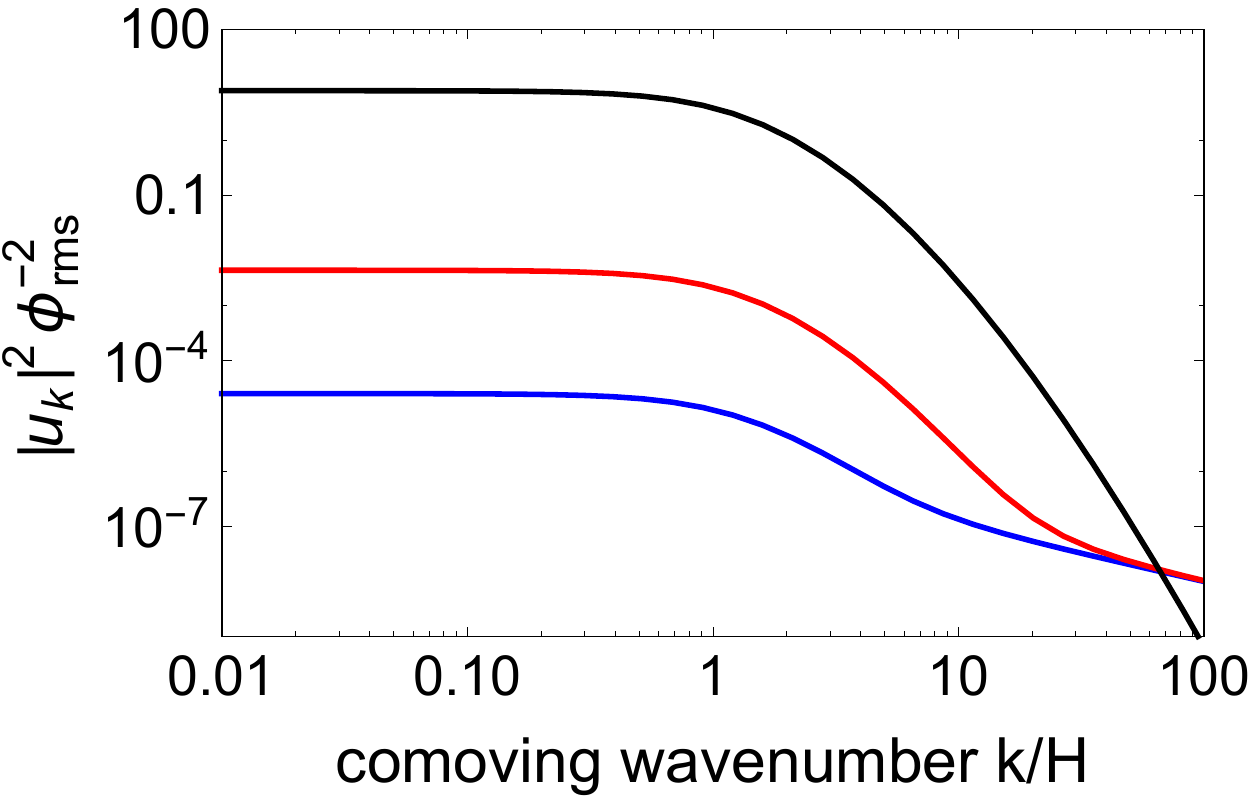}
\caption{
{\it Left:} The mode-functions for $k/H=0.1,1,10,100$ (blue, red-dashed, brown and green respectively) as a function of time in $e$-folds $N=Ht$. 
The black-dotted line shows the rms value $\phi_{\rm rms}^2$.
The waterfall transition occurs at $N=0$
{\it Right: } The normalized power spectrum $|u_k|^2/\phi_{\rm rms}^2$ as a function of wavenumber for $1$, $2$ and $14$ $e$-folds after the waterfall transition. The exponential suppression of the high-$k$ modes is evident. }
\label{fig:ukmodes}
\end{figure}

Fig.~\ref{fig:ukmodes} shows the amplification of a few characteristic mode-functions $u_k$ as a function of time. We see that before the waterfall transition (which we take to occur at $t=0$), the mode-functions follow the Bunch-Davies vacuum, red-shifting as radiation in an expanding universe. After the waterfall transition, the smaller wave-numbers $k\ll H$ start growing, due to the emergence of a tachyonic frequency-squared. At later times, more wavenumbers start growing. Fig.~\ref{fig:ukmodes} shows the  spectrum of the mode-functions $|u_k|^2$, where we see that it is dominated by modes with $k\sim H$. 
\beq
\ddot u_k + 3H\dot u_k +\omega^2_k u_k=0
\, ,\quad
\omega^2_k ={k^2\over a^2} -m_0^2 (1-e^{-pt})
\eeq
For modes with $k\ll H$, the behavior close to the transition is non-trivial and is discussed in detail in \cite{Guth:2012we}. For larger wavenumbers, we can easily show the exponential suppression by taking the approximation $e^{-pt}\ll 1$. This leads to the effective frequency-squared
$
\omega^2_k \simeq{k^2\over a^2} -m_0^2
$

which becomes negative for $k/a<m_0$. Since we consider an exact de-Sitter expansion during hybrid inflation $a=e^{Ht}$, meaning that large wave-numbers start growing at $Ht\simeq \log(k/m_0)$. Before this point the evolution equation for the mode-function is dominated by the $k^2/a^2$ term, meaning that they effectively redshift as radiation. Since $|u_k|$ decays exponentially until $Ht\simeq \log(k/m_0)$, the high-$k$ part of the spectrum is exponentially suppressed. 

After the first few $e$-folds of the waterfall transition, the modes with $k\lesssim H$ will exhibit the same growth-rate and as it is shown in the left panel of Fig.~\ref{fig:ukmodes}, these modes will dominate the rms value of the field $\phi_{\rm rms}$.

Intuitively the $\phi$ field close to the waterfall transition can be described as a wavepacket with a typical spread given by the de-Sitter fluctuations $\langle \phi^2 \rangle  =H^2/(2\pi^2)$. If we consider that after the transition the field evolves as 
\beq
\phi_{\rm rms} = \phi_{\rm rms}^{\rm initial} e^{\int_0^t \lambda dt'}
\eeq
 with $ \phi_{\rm rms}^{\rm initial}  = H\sqrt{2}\pi$, the result matches with the numerically computed $\phi_{\rm rms}$ after the first few $e$-folds. We will thus use this semi-analytical approximation for the amplification of the gauge-fields by the rolling waterfall field.

\subsection*{Gauge Field Dynamics}

 The gauge field mode-functions in turn follow the linearized equation of motion
 \beq
 \ddot A_k^\pm + H\dot A_k^\pm + 
 \left (
 {k^2\over a^2} \pm {1\over \Lambda} {k\over a} \dot \phi_{\rm rms}
 \right )A+k^\pm
 \eeq
This is Eq.~\eqref{eq:Akeom}, where we only considered one waterfall field and correspondingly $\Lambda_i\to \Lambda$. This is merely a notational simplification, since the evolution of all waterfall fields is identical and thus the gauge field production will be dominated by the one with the smallest value of $\Lambda_i$. As shown for example in \cite{Adshead:2015pva, Adshead:2016iae}, the amplification of the gauge fields during the tachyonic regime can be captured by the WKB approximation. In fact, since $\dot\phi_{\rm rms}$ grows exponentially and the gauge field amplification is exponentially sensitive to the effective frequency, gauge field production will predominately take place near the end of the waterfall transition, where $\phi_{\rm rms}\sim M_{\rm Pl}$ and $\dot \phi_{\rm rms}\sim \lambda M_{\rm Pl} \sim H M_{\rm Pl}$.

\begin{figure}[h!]
\centering
\includegraphics[width=.45\textwidth]{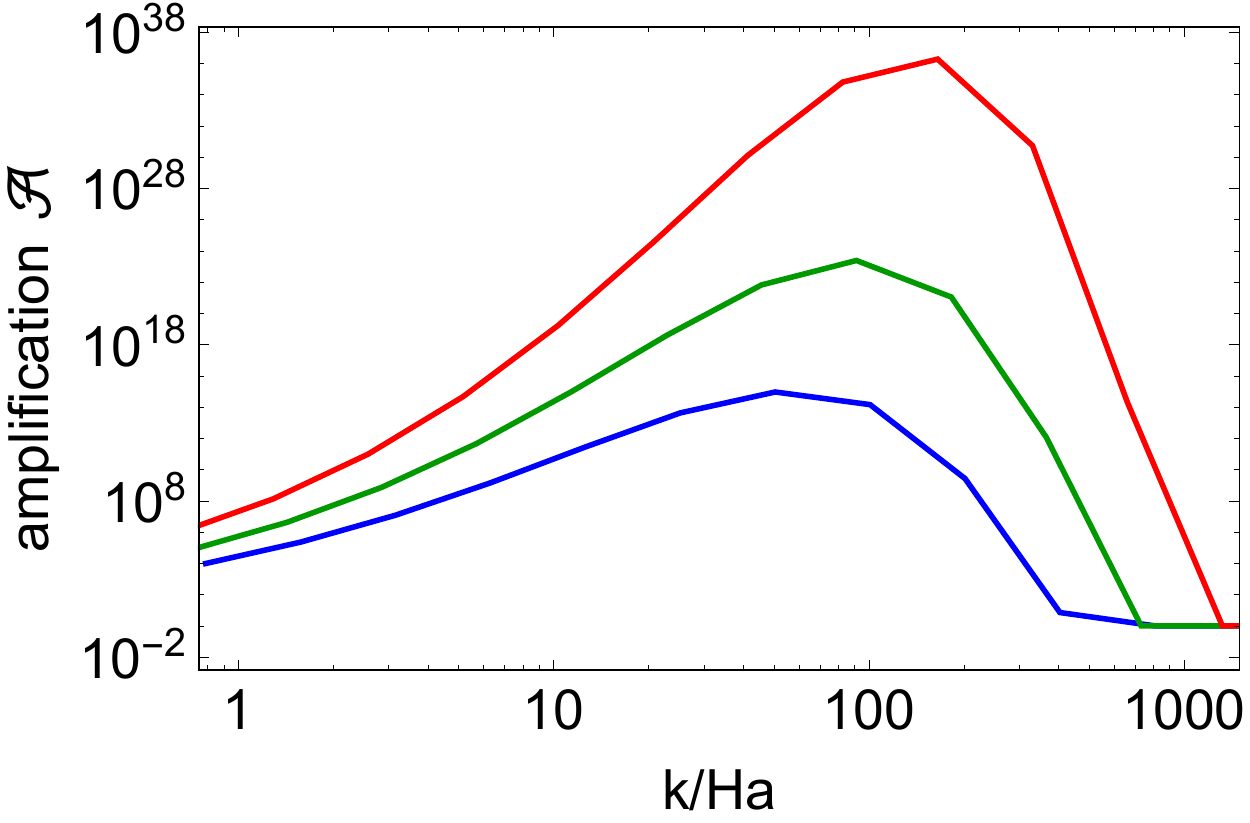}
\includegraphics[width=.45\textwidth]{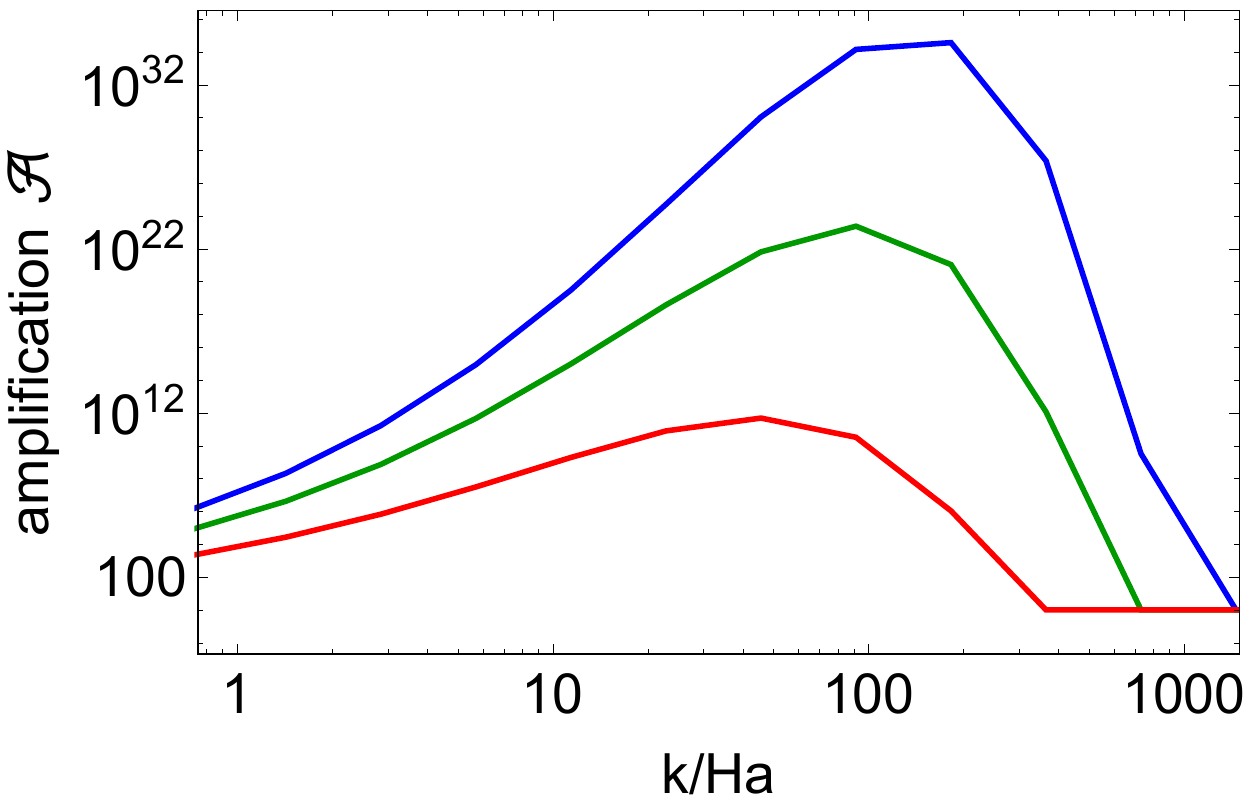}
\caption{
{\it Left:} The amplification factor ${\cal A}$ of the gauge fields for $14.2\pm0.1$ $e$-folds. 
{\it Right:} The amplification factor ${\cal A}$ of the gauge fields for $14.2$ $e$-folds and $\Lambda$ equal to the value shown in Table~\ref{tab1} increased or reduced by $50\%$.
We can see the very large dependence of ${\cal A}$ on the precise value of the number of waterfall fields and the strength of the Chern-Simmons coupling.
}
\label{fig:amplification}
\end{figure}

We must note the large sensitivity of the amplification to the exact value of the end-time of the waterfall phase and the exact value of $\Lambda$. Fig.~\ref{fig:amplification} shows the gauge field amplification 
for our benchmark example, as well as the corresponding amplification for small changes in $\Lambda$ and $t_{\rm end}$. 
In order for our calculations to be accurate, we need to maximize the energy transfer from the waterfall sector to the gauge field, while at the same time minimizing the back-reaction. This is why we require the two time-scales, the end of the waterfall phase and the completion of preheating, to coincide. This requires particular choices of $\Lambda_i$. Lattice simulations with the ability to capture the back-reaction effects will show the true parameter dependence of the model for smaller values of $\Lambda$ and hence stronger transfer of energy to the gauge field and thus an earlier onset of the strong-backreaction regime. We leave this numerical study for future work.

\section*{Comparison with existing literature}
Earlier work on GW's from hybrid inflation \cite{Garcia-Bellido:2007fiu, Garcia-Bellido:2007nns}
was restricted to a fast ending waterfall phase (less than one $e$-folds in order to avoid PBH production). In this case, the GW production mechanism is similar to a first order phase transition. The resulting GW's are weaker for lower frequencies,
leading to undetectable signals for interferometers in the near future, although could be within reach of BBO.

Single field models of axion inflation with a derivative (Chern-Simmons) coupling to abelian gauge fields have been shown to generate significant amounts of GW's with $\Omega_{\rm GW}\sim 10^{-8}$ \cite{Adshead:2019igv, Adshead:2019lbr}.
 While the coupling is similar to the one shown in Eq.~\eqref{eq:action}, single-field axion inflation in general produces signals at much higher frequencies, as shown in Refs.~\cite{Adshead:2019igv, Adshead:2019lbr} for a variety of potentials.
The difference from our model can be easily understood based on the estimate of Eq.~\eqref{eq:Omega}, where  $k_*/H\sim 10$ for single field (high-scale) axion inflation and  $k_*/H\sim 100$ for our model, leading to a factor of $100$ difference in $\Omega_{\rm GW}$.
However all single field axion models considered in \cite{Adshead:2019igv, Adshead:2019lbr} exhibit peak frequencies well above the interferometer range.
This arises from the structure of these single-field potentials, where the inflaton mass $m_\phi$ is comparable to the Hubble scale at the end of inflation. Producing the observed density perturbations requires $m_\phi \sim 10^{-6}\, M_{\rm Pl}$. 
This lead to high-frequency GW's with $\nu \gtrsim 10^6\, {\rm Hz}$. 
Furthermore, the GW amplitude does not depend on the peak frequency, both in Refs.~\cite{Adshead:2019igv, Adshead:2019lbr} and in our model.

Our model combines the two previous scenarios of axions and hybrid inflation. Specifically the waterfall transition in our case lasts for several $(\sim 10)$ $e$-folds, leading to a ``classical'' waterfall field motion. This sources the tachyonic resonance production of $U(1)$ gauge fields (like in single-field axion inflation~\cite{Adshead:2019igv, Adshead:2019lbr}), but operating at a much smaller Hubble scale. This allows for the production of detectable GW's in current and future experiments.


\end{document}